\documentclass[11pt]{article}
\usepackage[utf8]{inputenc}
\usepackage[a4paper, margin=0.79in]{geometry}
\usepackage{amsmath}
\usepackage{bm, upgreek}
\usepackage{blkarray}
\usepackage[hidelinks]{hyperref}
\usepackage{graphicx}
\usepackage{xcolor}
\definecolor{mylightgray}{RGB}{225, 225, 225}
\usepackage{lettrine}
\usepackage{lineno}
\usepackage[superscript]{cite}

\newcommand{\lr}[1]{\left(#1\right)}

\newcommand{\sfrac}[2]{^{#1}\!/\!_{#2}}

\renewenvironment{abstract}%
{\small\quotation\noindent\ignorespaces\vspace{-25pt}\paragraph*{Abstract.}}%
{\endquotation}

\title{The untapped power of a general theory of organismal metabolism}
\author{Marko Jusup$^1$ \& Michael R. Kearney$^2$}
\date{%
    $^1$Fisheries Resources Institute, Japan Fisheries Research and Education Agency, Yokohama 236-8648, Japan\\%
    $^2$School of BioSciences, The University of Melbourne, Melbourne, VIC 3010, Australia\\[2ex]%
}

\begin{document}

\maketitle

\begin{abstract}
What makes living things special is how they manage matter, energy, and entropy. A general theory of organismal metabolism should therefore be quantified in these three currencies while capturing the unique way they flow between individuals and their environments. We argue that such a theory has quietly arrived---`Dynamic Energy Budget' (DEB) theory---which conceptualises organisms as a series of macrochemical reactions that use energy to transform food into structured biomass and bioproducts while producing entropy. We show that such conceptualisation is deeply rooted in thermodynamic principles and that, with the help of a small set of biological assumptions, it underpins the emergence of fundamental ecophysiological phenomena, most notably the three-quarter power scaling of metabolism. Building on the subcellular nature of the theory, we unveil the eco-evolutionary relevance of coarse-graining biomass into qualitatively distinct, stoichiometricially fixed pools with implicitly regulated dynamics based on surface area-volume relations. We also show how generalised enzymes called `synthesising units' and an information-based state variable called `maturity' capture transitions between ecological and physiological metabolic interactions, and thereby transitions between unicellular and multicellular metabolic organisation. Formal theoretical frameworks make the constraints imposed by the laws of nature explicit, which in turn leads to better research hypotheses and avoids errors in reasoning. DEB theory uniquely applies thermodynamic formalism to organismal metabolism, linking biological processes across different scales through the transformation of matter and energy, the production of entropy, and the exchange of information. We propose ways in which the theory can inform trans-disciplinary efforts at the frontiers of the life sciences.
\end{abstract}

\vspace{18pt}

%\subsection*{Introduction}

%\linenumbers

\lettrine[lines=2, lraise=0.025, nindent=0em]{\textbf{T}}{}\lowercase{o} physicists, organisms are enigmatic. To biologists, physics is often to be avoided. Yet what is special about organisms must ultimately be explained and quantified within the frameworks provided by physics and chemistry. A general theory of living beings should be founded on a quantification of the dynamics of mass, energy, and entropy flows associated with the metabolism of whole organisms across their ontogeny, together with the role of information in these flows. The last century has seen massive advances in our understanding of the role of genetic information in biology\cite{davies2013, adami2004, strait1996} and new ideas are emerging about how organisms and cells use and share information during ontogeny\cite{levin2019}. But to fully capture organisms through the currencies of mass, energy, and entropy, we require a suitable model of metabolism. Over the last few decades, `Dynamic Energy Budget' (DEB) theory has emerged as a unified formulation of metabolic transformations that links all three currencies\cite{jusup2017, kooijman2010, sousa2008, kooijman2001}. We contend, however, that the theory has remained poorly known to researchers spanning physics and life sciences, and thus its potential to unite biology with the explanatory and predictive power of thermodynamics remains unrealised. This is problematic because without the rigour of a formal theoretical framework, research is fragmented, results biased, replicability low, generalisation poor, and effort wasted. The need for theory is now being widely recognised, and clarified, across biological sciences\cite{marquet2014, scheiner2010, roughgarden2009, scheiner2008, keller2000} and their social\cite{muthukrishna2019} `cousins'.

We present DEB theory with the aim to convey its generality and scope while separating the aspects that simply follow from thermodynamic principles and those requiring biologically explicit assumptions. This metabolic theory takes three innovative approaches to characterise organisms as thermodynamic systems: (i) coarse-graining microscale biological structures into stoichiometrically-fixed pools of macromolecules, (ii) conceptualising an organism as a composite of at least two qualitatively distinct macromolecular pools, and (iii) introducing a microscale mechanism for emergent, implicit regulation of relative pool dynamics. We emphasise the physical and pragmatic necessities, and evolutionary advantages, of these innovations. We discuss a range of practical and conceptual insights illustrative of the theory's potential place at the foundations of life sciences, including resolution of the long-standing puzzle of approximately $\sfrac{3}{4}$ scaling of metabolic rate with body mass, and how new metabolic entities form through transitions from ecological to physiological relationships.

\begin{figure}[!ht]
\noindent
\colorbox{mylightgray}
{
\begin{minipage}{0.973\textwidth}
\textbf{Box~1}. Key thermodynamic concepts and relationships.

\vspace{6pt}

\textbf{Thermodynamic system}: a bounded region of space characterised by macroscopic thermodynamic quantities like temperature, pressure, and volume. These macroscopic quantities emerge from the system's microscopic properties, classically represented by particle positions and momenta. Statistical mechanics treats positions and momenta as random variables to account for our ignorance regarding the microscopic state of a vast number of particles given a complete macroscopic description of the system.

\textbf{Internal energy}: a bulk property of thermodynamic systems caused by their constituents moving about and binding to one another.

\textbf{First law of thermodynamics}: a statement of energy conservation for thermodynamic systems that relates changes in a system's bulk (internal) energy with exchanges across its boundary (work, heat, and substances). Symbolically:
\begin{equation}
\mathrm{d}U = \updelta W + \updelta Q + \overline{\mathbf{h}}^\mathrm{T}\mathrm{d}\mathbf{M},
\label{eq:1stlaw}
\end{equation}
where $\mathrm{d}U$ denotes internal energy change, $\updelta W$ work, $\updelta Q$ heat exchange, and $\mathrm{d}\mathbf{M}$ substance exchange. Infinitesimal changes $\mathrm{d}U$ and $\mathrm{d}\mathbf{M}$ depend only on the initial and final states, while small quantities $\updelta W$ and $\updelta Q$ depend also on the specific path between those states. Negative $\updelta W$ and $\updelta Q$ decrease internal energy, indicating work performed by the system and heat outflow, respectively. The amounts of substances must be accompanied by a column-vector of conversion coefficients $\overline{\mathbf{h}}$, called molar enthalpy, to have the dimension of energy.

\textbf{Entropy}: often described as a measure of microscopic `disorder' or `information content' of a thermodynamic system. Specifically, entropy quantifies the delocalization of the joint position-momentum distribution of particles given the observed macroscopic state. A perfectly localised distribution, that is, an idealisation corresponding to a delta-function probability density and non-equilibrium conditions, indicates low entropy and requires few bits of information to describe the system. Conversely, a perfectly delocalized distribution, corresponding to a uniform probability density and equilibrium conditions, indicates high entropy and requires many bits of information to describe the system.

\textbf{Second law of thermodynamics}: a relationship between entropy and heat and, through the prism of statistical mechanics, between heat and information. The addition of heat makes the probability density of microscopic states more uniform, thus increasing entropy and information content. Symbolically:
\begin{equation}
\mathrm{d}S = \frac{\updelta Q}{T} + \updelta\sigma + \overline{\mathbf{s}}^\mathrm{T}\mathrm{d}\mathbf{M},
\end{equation}
where $\mathrm{d}S$ denotes entropy change, $\updelta Q$ heat exchange, $T$ the system's temperature, $\updelta\sigma$ entropy production (see below), $\mathrm{d}\mathbf{M}$ substance exchanges, and $\overline{\mathbf{s}}$ is a row-vector of conversion coefficients analogous to $\overline{\mathbf{h}}$, called molar entropy.

\textbf{Entropy production}: a measure of departure from equilibrium. It holds that $\updelta\sigma=0$ for reversible and near-equilibrium (i.e., quasistatic) irreversible processes without chemical reactions and $\updelta\sigma>0$ otherwise\cite{schmidtrohr2014}. A convenient way to quantify entropy production is by combining the first and second laws into:
\begin{equation}
T\updelta\sigma = \updelta W' - \mathrm{d}G + \overline{\bm{\upmu}}^\mathrm{T}\mathrm{d}\mathbf{M},
\label{eq:3rdlaw}
\end{equation}
where $\updelta W' = \updelta W + p\mathrm{d}V$ is non-expansionary work, $G=U+pV–TS$ is Gibbs free energy, and $\overline{\mu} = \overline{h} - T\overline{s}$ is chemical potential.

\end{minipage}
}
\end{figure}

\subsection*{Using pools to define organisms as thermodynamic systems}

\begin{figure}[!t]
\centering
\includegraphics[scale=1.0]{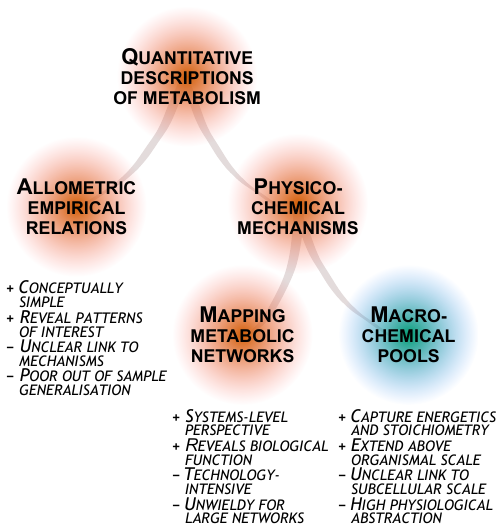}
\caption{Towards coarse-grained depictions of metabolism. Quantitative approaches in metabolic ecology have broadly followed two avenues: (i) seeking allometric empirical relationships or (ii) understanding physico-chemical mechanisms. For the latter, researchers typically adopt a microscopic perspective of mapping metabolic networks in detail. Alternatively, a macroscopic perspective can be taken, assigning metabolic activity to distinct molecular pools, each with a specific function. This macroscopic perspective strikes a balance between linking to underlying mechanisms and avoiding undue complexity.}
\label{fig:tree}
\end{figure}

Living organisms obey the laws of thermodynamics, which represent a coarse-grained description of processes operating at the microscopic scale (Box~1). Any thermodynamically sound metabolic theory must therefore start by coarse-graining the microscale biological structure and function (Figure~\ref{fig:tree}). The first innovation of DEB theory is to achieve this by conceiving of the organism as a set of stoichiometrically-fixed macromolecular pools undergoing chemical transformations. The formal application of thermodynamics thereafter leads to a quantitative framework governing organismal matter, energy, and entropy flows that is free of explicit biological assumptions, and which can act as constraints on evolutionary change. The framework is remarkably simple yet pertinent to complex organisms. It possesses substantial explanatory and predictive powers, making it an ideal tool for problem solving in biology.

Operating as non-equilibrium thermodynamic systems, living organisms actively maintain a steady state far from equilibrium\cite{vonbertalanffy1950}. This requires the acquisition and conversion of low-entropy food into cellular building blocks and energy through entropy-producing chemical reactions. Accordingly, building a formalised model of metabolism in DEB theory begins with specifying the number of macromolecular pools and the chemical transformations between them. The question is, however, just how coarse can the model representation of the organism be to balance the tradeoff between utility and complexity? Surprisingly, the answer is `quite'. But one pool is not enough.

\subsection*{The power of two: the need for qualitatively distinct metabolic pools}

Metabolic theories since Pütter\cite{kearney2021, vonbertalanffy1957} have implicitly represented organisms as a single macromolecular pool. The oversimplification of representing organisms as the chemical transformation of food into a single pool of biomass is made clear by the usefulness of the Fulton condition factor\cite{nash2006}---a ratio of body mass to volume. In a single-pool representation, such a ratio is a constant, a mere unit-conversion coefficient devoid of any additional informational value.

\begin{figure}[p]
\centering
\includegraphics[scale=1.0]{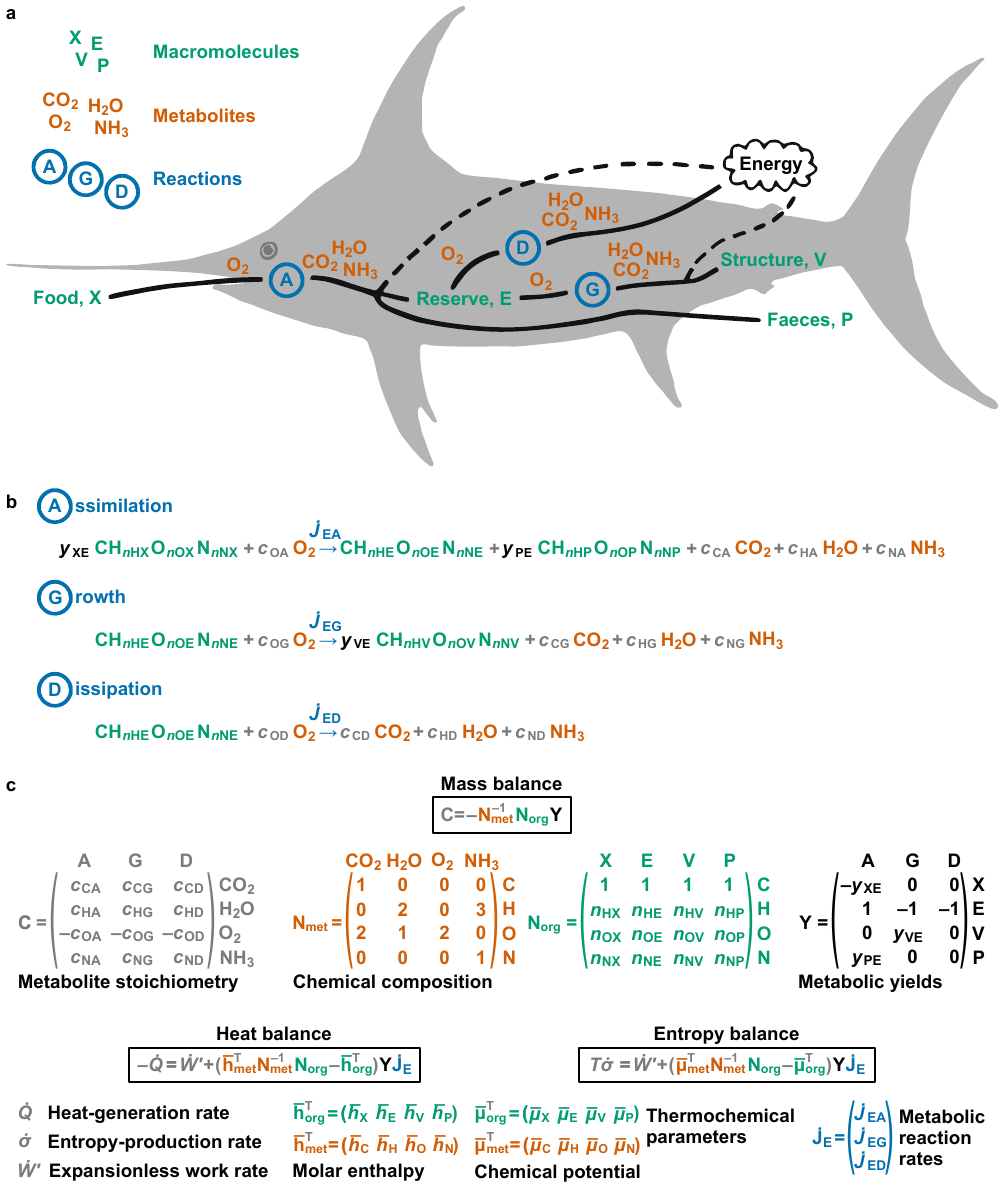}
\caption{Organisms as aggregations of macromolecular pools. (a) In a dual-pool representation, `reserve' E comprises metabolisable material accumulated by assimilating food X, and utilised to grow and maintain metabolically more inert `structure' V. Faeces P are produced parallel to assimilation as a result of imperfect digestion. (b) A two-pool organism performs three macrochemical reactions, exemplified here for heterotrophic aerobes whose nitrogenous waste is ammonia. Symbolically, X\,$\rightarrow$\,E\,$+$\,P is assimilation, E\,$\rightarrow$\,V is growth, and E\,$\rightarrow$\,energy is dissipation. (c) Applying the laws of thermodynamics to the macrochemical reactions yields the organism's heat generation rate $\dot{Q}$ and entropy-production rate $\dot{\sigma}$ prior to invoking any biological assumptions. See Supplementary material~1--3 for mathematical derivations.}
\label{fig:budgets}
\end{figure}

The single-pool representation also implicitly assumes a constant chemical composition within the organism. This becomes clear when considering that predicting a variable ratio of, for instance, proteins to lipids, would necessitate both the total biomass equation and an equation for the protein-to-lipid ratio, in effect subdividing the biomass into distinct protein and lipid pools. The macromolecular pools of DEB theory naturally incorporate changes in chemical composition, which are fundamental to ecological stoichiometry\cite{hessen2013} and stable isotope analysis\cite{west2006}. The latter in particular has garnered much attention by enabling the measurement of stable isotope ratios to trace nutrient cycles, food web dynamics, and animal migration patterns. Isotope dynamics emerging from DEB theory effectively capture important patterns revealed by stable isotope ratios\cite{pecquerie2010}.

Thus, the second innovation of DEB theory is to represent an organism as an aggregation of at least two qualitatively distinct macromolecular pools (Figure~\ref{fig:budgets}a), a metabolisable reserve (pool 1) accumulated through the assimilation of food and mobilised to build a metabolically more inert structure (pool 2). Reserve is a maintenance-free source of energy and materials that is replenished through feeding, whereas structure is not only built but is also maintained by mobilising reserve. A high level of order (and lower entropy) of structure requires energy and materials for maintenance.

The dual-pool representation must incorporate macrochemical transformations that account for the buildup of reserve and structure, as well as for the general energetic needs of organisms. Three such transformations are sufficient: assimilation, growth, and dissipation (Figure~\ref{fig:budgets}b). The assimilation reaction replenishes reserve from food. Faeces are produced in the process due to limits on digestibility, as reflected in the widely familiar concepts of digestible and metabolisable energy. Reserve is then used in the growth reaction to build structure, and in the dissipation reaction to provide metabolic energy. The assimilation and growth reactions primarily repurpose the materials in food and reserve, but some of those materials are also broken down, resulting in overheads that power the associated metabolic machinery. All three macrochemical transformations thus dissipate at least some energy from which emerges the `metabolic rate'.

Key quantitative results follow from mass-balancing the three macrochemical transformations, and subsequently applying equations \eqref{eq:1stlaw} and \eqref{eq:3rdlaw}. In our example (Figure~\ref{fig:budgets}c), we focus on the four primary elements---carbon C, hydrogen H, oxygen O, and nitrogen N---that collectively comprise 99\,\% of living biomass. Correspondingly, there are four primary `metabolites': carbon dioxide CO$_2$, water H$_2$O, oxygen O$_2$, and ammonia NH$_3$ (replaceable with other forms of nitrogenous waste such as urea, CH$_4$ON$_2$). The stoichiometric coefficients for these metabolites are collected in matrix $\mathbf{C}$, which is determined by the organism's metabolite composition $\mathbf{N}_\mathrm{met}$, organic-compound composition $\mathbf{N}_\mathrm{org}$, and metabolic yields $\mathbf{Y}$, the latter being constrained by biochemical rather than physical factors.

To further characterise metabolism, yields $\mathbf{Y}$ must be complemented with the rates of macrochemical transformations, collected in vector $\dot{\mathbf{J}}_\mathrm{E}$. Equation \eqref{eq:1stlaw} then leads to the organism's heat balance (Figure~\ref{fig:budgets}c) through a two-step process that involves taking the difference between energy flows due to substance outputs and inputs, and equating this difference with energy flows responsible for constructing reserve and structure. A similar rationale in conjunction with equation \eqref{eq:3rdlaw} leads to the organism's entropy balance (Figure~\ref{fig:budgets}c). This brings us to heat and entropy balances for individual organisms without invoking any biological assumptions. The argument rests entirely on thermodynamics with hints of organic chemistry and thermochemistry, as reflected in the choice of C, H, O, and N as the four primary elements, and the appearance of the thermochemical parameters in vectors $\overline{\mathbf{h}}_\mathrm{met}$, $\overline{\mathbf{h}}_\mathrm{org}$, $\overline{\bm{\upmu}}_\mathrm{met}$, and $\overline{\bm{\upmu}}_\mathrm{org}$.

\subsection*{From metabolic pools to metabolic rate and its scaling}

The metabolic rate and how it scales with body size is a fundamental ecophysiological phenomenon that emerges naturally from the dual-pool representation of organismal metabolism. By definition, the four metabolites are exchanged with the environment at rates $\dot{\mathbf{J}}_\mathrm{met} = \mathbf{C}\dot{\mathbf{J}}_\mathrm{E}$ (Figure~\ref{fig:budgets}c). This relationship is non-invertible in its full form, but it can be made so by focusing only on inhaled O$_2$ and exhaled CO$_2$ while recognising that growth is in many situations a slow process. The need for rows H$_2$O and NH$_3$, as well as the growth column G is thus eliminated from matrix $\mathbf{C}$, as is the need for growth rate $\dot{\mathbf{J}}_\mathrm{EG}$ from vector $\dot{\mathbf{J}}_\mathrm{E}$. We consequently get $\dot{\mathbf{J}}_\mathrm{E} = \mathbf{C}^{-1}\dot{\mathbf{J}}_\mathrm{met}$, which can be inserted into the heat-balance equation, showing that measurement of the exchange of two gases with the environment suffices to obtain an organism's heat-generation rate. This empirical fact, foundational in clinical applications of indirect calorimetry\cite{ferrannini1988}, here finds theoretical explanation and support.

\begin{figure}[!t]
\centering
\includegraphics[scale=1.0]{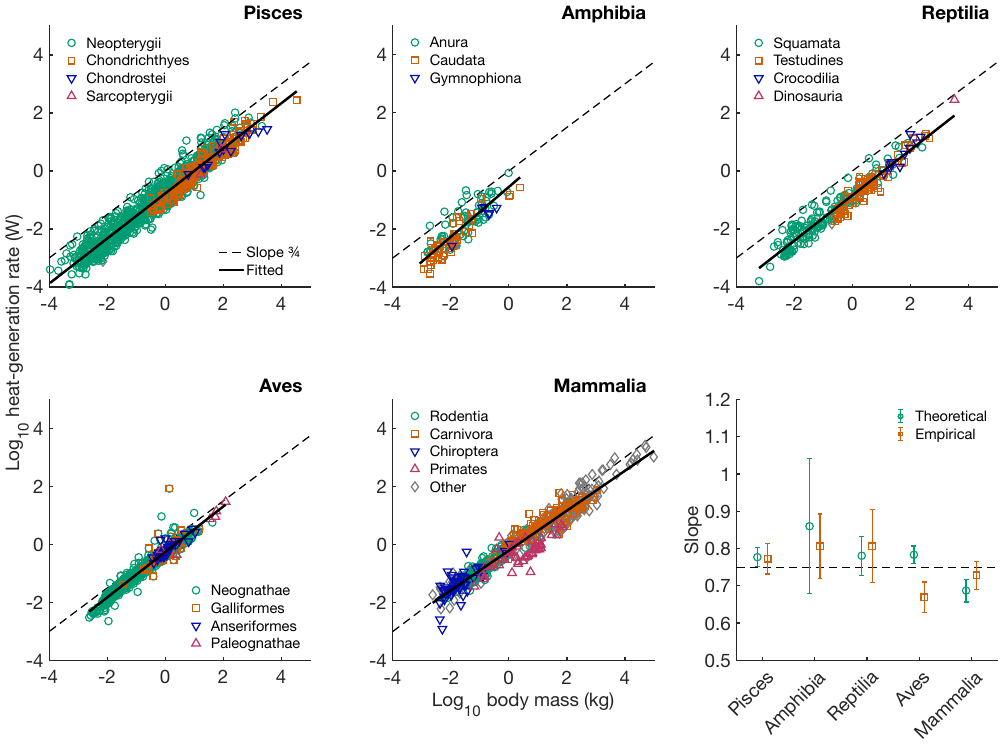}
\caption{From theory to iconic patterns of life; metabolic body-size scaling in vertebrates. The dual-pool representation of organismal metabolism makes no explicit assumptions related to the approximate $\sfrac{3}{4}$ scaling exponent that is frequently observed when a metabolic rate is plotted against body mass for species across a wide range. And yet, the theory predicts the familiar scaling law due to Kleiber with sufficient fidelity to capture the variation between taxonomic classes. Among vertebrates, only the prediction for birds differs significantly from documented empirical values\cite{white2019}. The data and code to generate this figure are available online\cite{osf2024}.}
\label{fig:scaling}
\end{figure}

The scaling of heat dissipation with size, both through ontogeny and between species, is empirically observed to be a power function of body mass where the exponent varies but is frequently near $\sfrac{3}{4}$\cite{dodds2001, kleiber1947, kleiber1932}. Much thought has been invested in the underlying causes of this taxonomically general empirical pattern\cite{white2022, white2019, glazier2010, vandermeer2006, brown2004}, but it is regarded as unsolved. The dual-pool thermodynamic processes of DEB theory lead to metabolic rate being a weighted sum of the heat lost through assimilation, growth, and dissipation. That this should produce a scaling of metabolic rate with an exponent of roughly $\sfrac{3}{4}$ was pointed out nearly 40 years ago in the origins of DEB theory\cite{kooijman1986}. DEB parameter estimates are now available for thousands of vertebrate species\cite{kooijman2021}. Most of these parameter estimates were based only on observations of lengths and weights at key life-cycle stages as well as reproduction and development rate. We explicitly excluded (rare) datasets on respiration, and yet the estimated parameters imply respiration rates and scalings that match strikingly with observation (Figure~\ref{fig:scaling}). The dual-pool representation lacks any explicit biological assumptions that would account for such scaling and, unlike contending explanations \cite{white2022, west1997}, uses no optimality arguments. Instead, scaling emerges from physical relations of surface area and volume between reserve and structure, together with the fact that structure requires maintenance but reserve does not\cite{maino2014}. The microscale origins of such dynamics are discussed next.

\subsection*{Pools but not swimming pools: structured chemical reactions with implicit metabolic regulation}

A century ago, Lotka had the insight that living things will need to be characterised by structured chemical reactions instead of homogeneous `swimming pool' chemistry: ``Physical chemistry is still a comparatively young science, and naturally the simpler phenomena have been sought out for first attention. This is not because complex physico-chemical structures do not exist, nor even because they are unimportant. On the contrary, it is to be expected that the future will bring important developments''\cite{lotka1925}. Lotka emphasised that in structured chemical dynamics the geometrical features will play a dominant role because different structures must grow together in such a way as to maintain suitable environments for each other. That is, as the reaction proceeds, the parts mutually construct new environments that must be compatible with their persistence.

\begin{figure}[!t]
\centering
\includegraphics[scale=1.0]{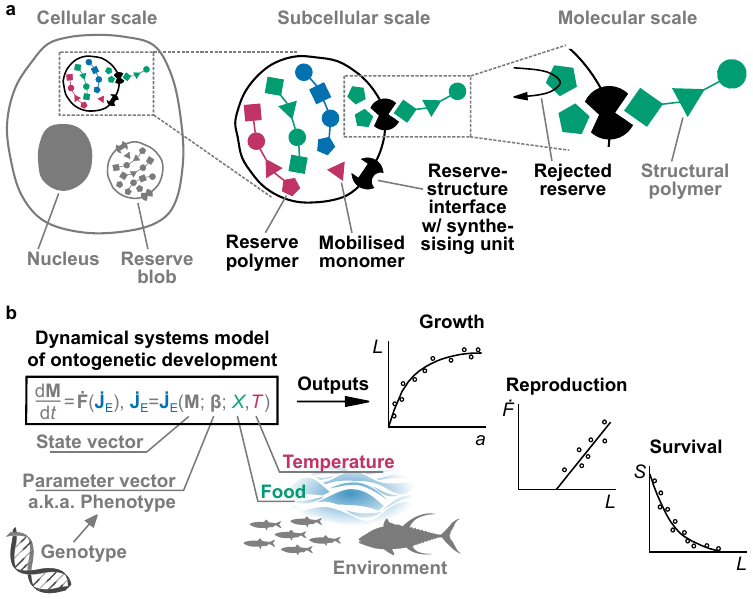}
\caption{Bridging physics and biology via dynamical systems models of ontogenetic development. (a) In DEB theory, the relative dynamics of reserve to structure is resolved by having the reserve stored as blobs of polymers to avoid osmotic problems. Reserve and structure then interact through their mutual surface-area interface via stylised enzymes called `synthesising units'. These enzymes, bound to the reserve-structure interface, act depending on the delivery rate of reserve material to them and the processing time for that material. (b) Predicting mass, heat, and entropy balances throughout ontogeny necessitates expressing metabolic reaction rates $\dot{\mathbf{J}}_\mathrm{E}$ as functions of the organism's state $\mathbf{M}$, alongside phenotypic parameters $\mathrm{\upbeta}$ and environmental drivers (e.g., food $X$ and temperature $T$). The time evolution of $\mathbf{M}$ is given by a vector function \smash{$\dot{\mathrm{F}}=\dot{\mathrm{F}}\lr{\dot{\mathbf{J}}_\mathrm{E}}$} whose reserve and structural components are defined by the dual-pool representation of metabolism, but which may also incorporate additional components (e.g., maturation status, reproductive investments, cellular damage, etc.). See Supplementary material~5 for a DEB-based dynamical systems model of ontogeny.}
\label{fig:dsm}
\end{figure}

The third innovation of DEB theory captures Lotka's foresight by characterising metabolism as structured chemical transformations that are regulated implicitly by geometrical relationships. The reasoning begins with the fact that reserve material must be stored as blobs of polymers to avoid osmotic problems, and therefore must be accessed via the surface-area interface with structure at the subcellular level (Figure~\ref{fig:dsm}a). This gives rise to fundamentally different dynamics from the law of mass action in well-mixed pools of monomers. Such dynamics can be captured with the concept of `synthesising units', which are stylised enzymes with a busy period when bound to substrates\cite{muller2019, kooijman1998}. These enzymes possess a strong affinity for their substrate, meaning that the transformations they catalyse are driven by substrate arrival rates (i.e., not by substrate concentrations) and product dissociation rates (i.e., substrate dissociation rates are zero). A rich array of structured chemical transformations can be characterised with synthesising units, including complementary and substitutable substrates transformed in series or in parallel, in chains via `handshaking' protocols, with preference or inhibition, and with smooth transitions between limiting substrates (avoiding the need for switches in mathematical formulations).

Substrate arrival and product dissociation jointly determine the relative dynamics of reserve and structure. Substrate arrival for synthesising units bound to reserve's surface-area interface with structure is a property of reserve, characterised solely by the reserve turnover rate under a given feeding regimen. This rate reflects the geometry of the reserve-structure interface. In isomorphic cells, for example, surface areas decrease relative to volumes inversely proportional to a characteristic volumetric length. The same holds for the reserve turnover rate because the growing reserve volume gradually overshadows its `access window', that is, the surface area of the reserve-structure interface. This puts a limit on the size of isomorphic growers when the reserve-structure interface becomes a bottleneck for satisfying the maintenance needs of a growing structural volume.

In contrast to substrate arrival, product dissociation is a property of synthesising units. A synthesising unit that is occupied processing the substrate (i.e., before products can dissociate) necessarily rejects any additional inbound mobilised reserve. Synthesising units are thus collectively responsible for a flow of rejected material that is returned to reserve, and that influences the reserve growth rate. Remarkably, there is a precise number of synthesising units per amount of structure for which the growth rates of reserve and structure balance one another\cite{kooijman2007}. The consequence is that a stable feeding regimen also stabilises the ratio of reserve to structure, called reserve density (the `weak homeostasis' assumption of DEB theory).

The ability to stabilise reserve density greatly simplifies the regulation of organismal chemistry. Namely, a change in the feeding regimen is necessary for assimilation to either temporarily outpace mobilisation or vice versa. A consequent gain or loss of reserve density upregulates or downregulates mobilisation in accordance with assimilation under the new feeding regimen. Structural growth follows suit, but so does the flow of rejected reserve material and ultimately the growth of reserve. Reserve and structure therefore cannot outgrow one another, nor can they wildly fluctuate relative to one another, and this occurs without a need to track the chemical inventory of cells or to deploy an explicit regulatory mechanism. The parts therefore mutually construct a stable internal environment as they grow, as Lotka emphasised.

The described `implicit regulation' of reserve density is broadly compatible with observations. This is evident from comparing the resulting dynamics of reserve density with the known facts about Fulton's condition. Specifically, reserve density closely tracks environmental food abundance, including staying steady when food abundance is steady. Likewise, changes in Fulton's condition generally follow changes in food abundance\cite{utne2021, jusup2014, abdeltawwab2006, carruthers2005, booth1999}, while the condition is stable under a stable feeding regimen. Further support comes from cases of sexual dimorphism in mass across the animal kingdom\cite{lika2024}. The largest sex is heavier for the same length because the structural mass is the same, but the reserve mass is not.

The implicit regulation of reserve density dynamics issuing from the `weak homeostasis' assumption of DEB theory has both biological and practical significance. The biological significance is that it saves energy and creates evolutionary flexibility. Energy is saved because no bookkeeping, and hence costly information manipulation\cite{fields2021, kempes2017, mehta2012}, is required to regulate the system. Evolutionary flexibility comes from the `mergeability' and `partitionability' of reserve that, in turn, allows varying degrees of mutual syntrophy with the extreme case being the evolution of the eukaryotic cell (see below).

The practical significance of the implicit regulation of reserve dynamics is that it renders the dual-pool representation tractable with a dynamical systems model. This is because model parameter estimates can be obtained from a small minimum set of observations of lengths and body masses, together with reproduction and development rates, at different stages of the life cycle\cite{kooijman2008}. Such a model then enables inference of mass, energy, and entropy balances throughout ontogeny by capturing the interplay between metabolic reaction rates and state variables (Figure~\ref{fig:dsm}b). Metabolic reaction rates govern the temporal change of state variables, while being functionally dependent on these variables (alongside phenotypic parameters and environmental drivers). The concrete form of the functional dependence is obtained by taking a leap from physics into the realm of biology, for which DEB theory offers a promising way forward (see Supplementary material~5).

\subsection*{The need for entropy production}

The concept of entropy has proven challenging to apply in biology, especially its relation to biological organisation and the distinction between thermodynamic and statistical interpretations\cite{berry1995}. DEB theory permits the calculation of an organism's total entropy and its production of entropy, in the thermodynamic sense, irrespective of its state (growing, starving, or reproducing) and underlying chemistry (aerobic or anaerobic). This breakthrough comes directly from conceiving biomass as stoichiometrically fixed pools, and subsequently applying the formalism of thermodynamics to the between-pool chemical transformations. But how is this useful?

One conceptual advantage is in clarifying the distinction between entropy and organisation. The quantification of the entropy of the whole organism depends on its reserve density but the entropy of structure, which defines biological organisation, is independent of the reserve density. Such a perspective resolves the paradox of a substrate having a lower entropy than the biomass growing on that substrate\cite{battley1992}. Namely, it is a given amount of structure, not biomass, that should (and always will) have a lower entropy than the same amount of substrate\cite{sousa2006}.

Equation~\eqref{eq:3rdlaw} and its DEB-based reformulation in Figure~\ref{fig:budgets} are conceptually and practically useful because they quantitatively tie feeding and movement to survival, growth, and reproduction in a strict thermodynamic sense. If we take it as given that an organism exists, then the thermodynamics tells us that this non-equilibrium system can only persist through the production of a defined amount of entropy, $T\updelta\sigma$. The production of entropy at this fixed rate defines the `existential imperative' of the organism, a concept being used to explore problems in agency, mentality, and intelligence in natural and artificial contexts\cite{ramstead2020}. Equation~\eqref{eq:3rdlaw} makes clear that, physically, the imperative must be resolved by making the net useful energy input from food, $\overline{\bm{\upmu}}^\mathrm{T}\mathrm{d}\mathbf{M}$, the same magnitude as the entropy production $T\updelta\sigma$ plus any work $W'$, primarily movement, needed to find this food and avoid hazards, and any increase required in Gibbs free energy, $\mathrm{d}G$, necessary for growth and reproduction ($\mathrm{d}G$ relates to the internal state and is negative when the organism is starving and positive when it is growing or reproducing). The quantification of these relations provides a rigorous thermodynamic basis for `mechanistic niche models' of whether organisms can persist under different environments\cite{kearney2009}.

Most entropy production is dissipated in the form of heat resulting in what is empirically described as the metabolic rate of organisms. Practically, metabolic heat generation is needed to solve biophysical problems such as water vapour exchange\cite{kearney2023} and heat stress\cite{mckechnie2010} that tie into the question of environmental limits. Metabolic heat production is a fundamental measure in comparative physiology\cite{withers1992}. That DEB theory permits heat flux to be computed directly (Figure~\ref{fig:budgets}) obviates the need for the use of empirical approximations from allometric functions, and such a computation is now possible for thousands of animal species (Figure~\ref{fig:scaling}).

Finally, stress-induced entropy production from free-radicals and other chemical stressors is non-exportable\cite{toussaint1998}, acting akin to a ticking clock for individual organisms through cellular damage and ageing\cite{lenart2016}. The dual-pool representation of DEB theory offers a rigorous yet practical way of quantitatively linking metabolic activity to cellular damage and ageing hazard\cite{vanleeuwen2010}.

\begin{figure}[p]
\noindent
\colorbox{mylightgray}
{
\begin{minipage}{0.973\textwidth}
\textbf{Box~2}. Evolution of metabolic organization and homeostasis.

\vspace{6pt}

\includegraphics[width=\textwidth]{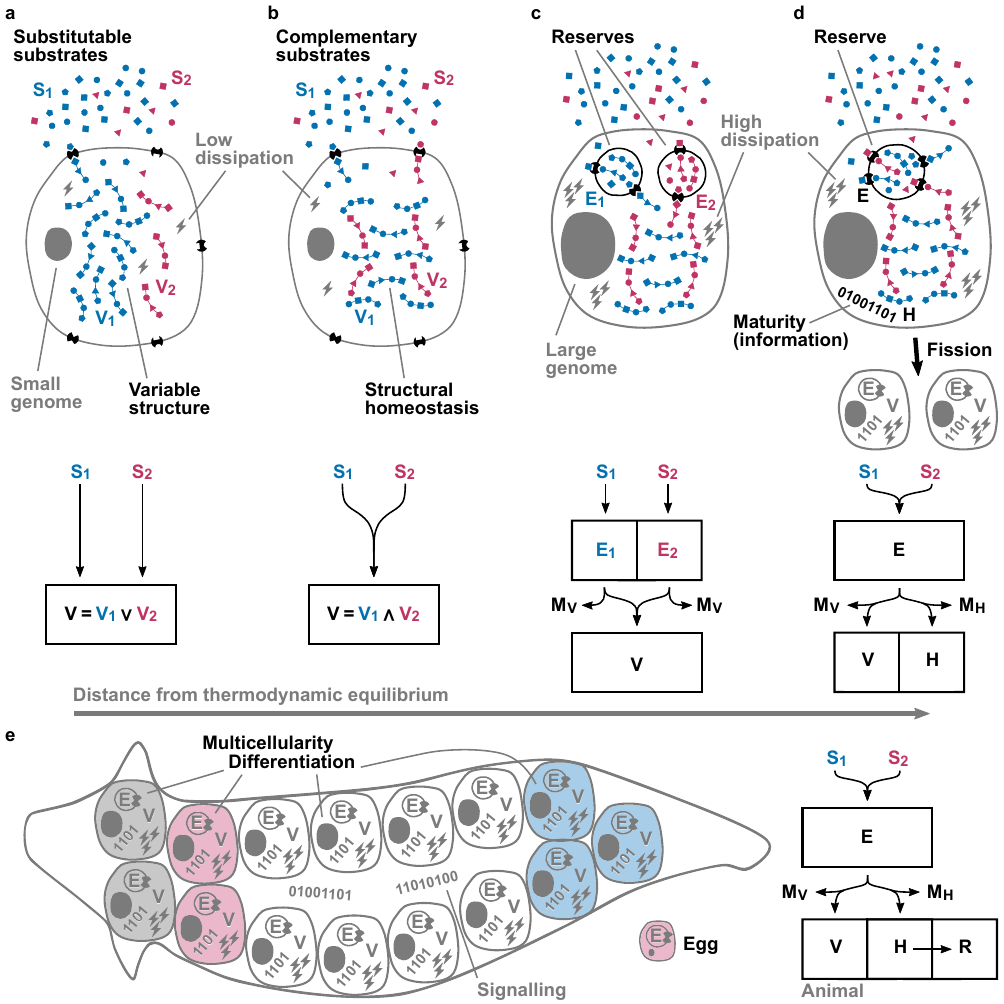}

From the perspective of Dynamic Energy Budget theory, a primitive non-equilibrium thermodynamic state resembling metabolic activity would have been reserveless and reliant on substitutable substrates to build and power stoichiometrically volatile structure (a). An initial step towards more structural homeostasis happens with the appearance of tightly coupled uptake of supplementary substrates (b). To ease the constraints on nutritional environment due to coupled uptake, substrates are pre-processed and internalised in reserve pools for subsequent use (c). Polymerization of reserve pools increases their storage capacity and reliability as internal energy sources. Greater reserve capacity also allows for the evolution of more complex structures, leading to higher maintenance and overhead costs, increased dissipation (metabolic rate), and larger genome sizes (d). Gene-regulation processes give rise to a cell cycle with binary fission, conditional on sufficient energy investment in maturation (i.e., metabolic information content modulated via gene regulation). The evolution of multicellularity drives differentiation and metabolic learning, increasing information content maintained in organisms, including paracrine, endocrine, and bioelectric cell-cell communication (e). The life cycle emerges when reserve, once maturation is completed, gets packaged with nuclear information in the form of eggs to restart the system.

\end{minipage}
}
\end{figure}

\subsection*{Entropy, evolution, and information}

Life thrives on the energy flows that occur where contrasting thermodynamic systems interact to equilibrate, such as hydrothermal vents and radiative exchange between the sun and the earth's surface. But to understand why in such places thermodynamics promotes the self-organisation of structurally complex (i.e., low-entropy) entities, it is necessary to focus on the production of entropy. The explicit status of entropy in DEB theory opens the door to new, quantitative perspectives on the connections between entropy, evolution, and information.

DEB theory makes clear that entropy production is the principle measure of metabolic rate; heat production is a close second but does not always generalise to anaerobic metabolism. Boltzmann and Lotka paved the way to a view that entropy production is central to evolution too, specifically through the `Maximum Entropy Production Principle'\cite{hall2023, vanchurin2022, kleidon2010, martyushev2006}. According to this principle, thermodynamic systems equilibrate as fast as the physical processes permit, which is facilitated by maintaining local structural complexity that increases the entropy of the surrounding space. Viewed through this prism, life's `structural complexity' emerges as an accumulation of homeostatic mechanisms that enhance metabolic stability\cite{kooijman2007}, while paying for such stability with ever increasing dissipation (i.e., entropy production).

DEB theory allows thermodynamic formalism to be applied to models of the evolution of metabolic organisation. The explicit quantification of entropy production is fundamental to this capacity, but so too is the dual-pool representation of the organism in terms of reserve and structure, together with the modularity that stems from the implicit regulation of reserve dynamics (Box~2). A gradual increase in structural complexity and chemical homeostasis happens as the uncoupled uptake of substitutable compounds transitions to the coupled uptake of supplementary compounds, and substrates get internalised into reserve pools (Box~2, panels~a--c). Moving towards single-reserve metabolism, thermal response curves for chemical transformations involved in different metabolic processes converge towards one another to avoid complex regulatory problems associated with temperature-sensitive variation in yields (Box~2, panels~d\,\&\,e). The result is more tightly integrated homeostatic systems, exemplified by animals. Thus, the simple one structure, one reserve system we have used to introduce the theory (Figure~\ref{fig:budgets}) is in fact a highly evolved metabolic state. In contrast to animals, even the most evolved terrestrial vascular plants cannot be reduced to less than a six-pool system in which roots and shoots each comprise one structure on top of two interconnected reserves\cite{russo2022, schouten2020}.

There is a growing appreciation that major evolutionary steps have occurred through the merging of existing systems to form new ones\cite{nanos2022}. With the transition from substitutable to supplementary nutrients, the reserve dynamics framework quantifies how non-limiting reserves can dam up. Excretion of these excess reserves to the environment, together with other types of metabolic product formation, then creates opportunities for mutual syntrophy (e.g., microbiomes\cite{marschmann2024}) and obligate syntrophic relationships (e.g., corals and lichens) that, in the extreme, involve the complete merging of metabolic systems (e.g., mitochondria and chloroplasts). Thus intra- or interspecies ecological relations can evolve to become physiological relations within new, composite individuals. Such transitions can be modelled using eco-metabolic formulations\cite{kooijman2003} that represent cases of mutual niche construction\cite{constant2018, odlingsmee2003}.

As the evolution of metabolic organisation generates entities further from thermodynamic equilibrium, those entities necessarily produce more entropy (dissipation) but must also process more information (learning). The dual-pool representation of DEB theory readily incorporates information processing through the state variable `maturity'. The formal dimension of maturity is information but, in practice, maturity is quantified as cumulated energy invested in the processes of maturation. The installation of the maturation process in unicellulars enables the tight coordination of the cell cycle, including gene regulation for membrane and cytoskeleton development, and DNA replication. With the transition to multicellularity, maturation processes comprise cell-cell communication to coordinate emergent phenomena such as cell differentiation, bioelectrical signalling\cite{levin2019a}, hormonal regulation, and immune systems. Maturity has a maintenance energetic cost proportional to the level of maturity attained; throughout the lifecycle, chemical and electrical patterns are maintained to retain developmental order in the face of perturbations. This maturity maintenance cost is in addition to the energy demanded for the maintenance of the structural component of the biomass.

The maturity state variable creates energetic hooks into the rapidly progressing field of developmental biology where the role of collectively held information among cells is becoming clearer\cite{levin2019b}. DEB theory makes distinct predictions for how growth and development may become decoupled depending on the nutritional and thermal environment, creating fertile ground for experiment. It also provides a place in the metabolic architecture to explicitly account for constraints and trade-offs on how energy is spent on cellular organisation and communication schemes necessary for multicellular development and learning.

\subsection*{From molecules to ecosystems: Bridging the scales of biological organisation}

Metabolic theories in ecology inherently focus on the individual scale. There is, however, significant potential to bridge scales from molecules to ecosystems\cite{nisbet2000}. Transitioning to higher scales---populations, communities, and ecosystems---is relatively straightforward because populations are aggregates of individuals. Predicting individual ontogeny under varying environmental conditions directly informs population dynamics, effectively making it a matter of bookkeeping\cite{persson2003, deroos1997}. Once populations interact, they form communities, where individual performance can cause surprising dynamics\cite{deroos2003, deroos2002}. Even the very stability of complex communities in large part depends on individual performance\cite{deroos2021}.

Transitioning to below-individual scales---cellular, subcellular, and molecular---is less straightforward. An established approach is to treat chemicals as another environmental driver (besides food, temperature, etc.) whose pharmacokinetics are tied to metabolic reaction rates (\textit{sensu} DEB), and whose accumulation or throughput affects the phenotypic parameters of the dynamical systems models of ontogeny\cite{jager2023, jager2020, klanjscek2007}. This method allows for rapid assessment of the full range of environmental interactions due to novel industrial compounds, such as engineered nanomaterials\cite{holden2013}, which may also cause surprising twists in mutualistic dynamics when individual-scale performance is accounted for\cite{klanjscek2017}.

Although tying pharmacokinetics to metabolic reaction rates predicts the fate of single molecules, it falls short of establishing the metabolic basis behind macromolecular pools and flows between them. Information on what constitutes reserve and structure may be attainable by mapping metabolic networks and analysing them using methods such as flux-balance analysis (FBA)\cite{kauffman2003}. FBA applies linear programming to determine steady-state reaction rates across a metabolic network, respecting stoichiometric and some thermodynamic constraints. It requires an objective function to optimise, such as biomass growth, ATP synthesis, or minimum nutrient usage. Given these characteristics, hybrid DEB-FBA models have the potential to inform each other, with the DEB part predicting the bioavailability of key chemicals, and the FBA part predicting their biochemical effects.

FBA is a powerful tool for metabolic engineering, but it also comes with substantial limitations. Some of these limitations are addressed within related constraint-based reconstruction and analysis (COBRA) approaches\cite{antoniewicz2021}; for example, metabolic-flux analysis eliminates the need for an objective function. Model selection and validation remain further practical concerns for COBRA approaches\cite{kaste2024}. For mechanistic understanding, however, the most limiting factors are the steady-state assumption and the omission of regulatory mechanisms. This is where further integration with kinetic and regulatory-network models can help. Kinetic models\cite{almquist2014} incorporate dynamic aspects of metabolic processes, providing insights into the temporal behaviour of metabolic reactions beyond the steady-state assumption. Regulatory-network models\cite{bocci2023} connect to genomics by mapping out gene regulatory mechanisms that control metabolic pathways, thus unveiling the genetic and epigenetic basis of the phenotypic parameters required by the dynamical systems model of ontogeny. When detailed mapping of metabolic and regulatory networks is unavailable, statistical relations between genome data and phenotypic parameters can be established to predict emergent behaviours\cite{marschmann2024}.

\subsection*{Conclusion}

As emphasised by Brown et al.\cite{brown2004} and Nisbet et al.\cite{nisbet2000}, a solid theory of individual metabolism could provide a powerful theoretical foundation for linking biological processes across different scales through the transformation of matter and energy, and the exchange of information, all within the constraints imposed by entropy (irreversibility and dissipation, steady-state maintenance, and useful work). Here, we aimed to show that DEB theory has the potential to take this role, and we see no viable alternative.

Much of what constitutes DEB theory is simply the careful application of basic thermodynamic principles to organisms. But the theory's power comes from important innovations in the setup (the dual-pool representation of metabolic transformations) and formulation (implicit regulation of reserve and synthesizing units), all of which pertain to the subcellular scale. That a general theory of metabolism focuses on subcellular mechanisms should not come as a surprise given the evolution's tendency to co-opt existing biochemical pathways for new purposes\cite{jeong2000}. Unlike the remarkable diversity of forms and functions observed across the tree of life starting with unicellular organisms onwards, the subcellular scale exhibits a surprising degree of biochemical uniformity. This is well exemplified by the second-messenger molecule cyclic AMP, which is found in archaea, bacteria, and eukaryotic cells of plants and animals alike, where it serves an array of functions ranging from signalling hunger in \textit{E. coli} to forming short-term memory in brains. DEB theory, with its emphasis on subcellular mechanisms and transformations, is well-suited to capture this fundamental unity amidst the stunning phenotypic diversity of life.

Biology faces a theory crisis, in large part due to a lack of formal theoretical frameworks\cite{marquet2014, scheiner2010}. The utility of such frameworks is multifaceted, but often stems from their explicit emphasis on the constraints imposed by the laws of nature, which improves hypothesis formulation and prevents reasoning errors. Otherwise, fragmented approaches may lead to incomplete or contradictory findings\cite{belovsky2004}, causing research waste\cite{purgar2022, grainger2020} and hindering scientific debate\cite{pauly2021}.

Organismal metabolism is at the heart of many open problems in ecophysiology and the life sciences, including our most pressing environmental and philosophical problems. Yet a thermodynamically formalised theoretical framework for studying metabolism has never been widely accepted. As we have shown, such a framework exists in the form of DEB theory, which offers a simple yet realistic and robust quantification of metabolism as the transformation of matter and energy, the production of entropy, and the exchange of information. By founding our questions about metabolism upon a thermodynamically grounded theoretical framework like DEB theory, or any alternative approaches if they are found to exist\cite{kooijman2020}, we will generate better questions, hypotheses and tests, make better predictions, and develop greater understanding.

%\nolinenumbers

\paragraph*{Acknowledgments} M.\,J. was supported by the Japan Society for the Promotion of Science (JSPS) KAKENHI grant no. 21H03625. M.\,R.\,K. was supported by Australian Research Council, Grant/Award Number: DP200101279. We thank Steven Chown, Tiago Domingos, Karl Friston, Bas Kooijman, Nina Marn, Roger Nisbet, and Tania Sousa for insightful feedback on the manuscript, and Craig White for providing confidence intervals for Figure 3.


\begin{thebibliography}{99}
\bibitem{roughgarden2009} Roughgarden J. Is there a general theory of community ecology? Biol Philos 24(4):521-9.
\end{thebibliography}
\end{document}

\maketitle

\subsection*{Introduction}

The purpose of this text is twofold: (i) to lay out a formal line of reasoning that gives rise to mass, heat, and entropy balances within the framework of the `Dynamic Energy Budget' (DEB) theory, and (ii) to outline the standard dynamical systems model of ontogeny according to DEB theory. It is assumed that the reader is familiar with thermodynamic principles listed in Box~1 of the main text. For convenience, all necessary DEB concepts and the corresponding mathematical notation will be defined here, even if the main text contains equivalent definitions.

\subsection{Mass balance}

An organism is assumed to comprise two macromolecular pools, reserve and structure. The former is metabolisable as a source of energy and building blocks for growth. The latter is more inert due to being critical for the organism's survival. Accordingly, substances acquired through ingestion and subsequent assimilation are temporarily stored in reserve, whose transitory nature needs negligible maintenance and is well characterised by a turnover time. Substances mobilised from reserve as building blocks are embedded more permanently in structure, where they get replaced to counter the inevitable degradation due to intrinsic disturbances such as oxidation. Structure therefore requires constant maintenance.

The organism, and thus reserve and structure, is assumed to occupy a certain volume in space, called the control volume. Associated with the control volume is the control surface that separates the organism from the environment. Control volume and surface are thermodynamic concepts typically introduced to track the exchange of substances and heat across the control surface, as well as the production of entropy inside the control volume.

The state of the organism, or formally, the state of the system inside the control volume is tracked via the amounts of substances in reserve and structure, $M_\mathrm{E}$ and $M_\mathrm{V}$, respectively. Reserve and structure have a different chemical composition from one another, but the composition of each individually is assumed to stay constant in time (the `strong homeostasis' assumption of DEB theory).

To build reserve and structure, the organism needs to exchange substances with the environment. The amount of ingested food is denoted $M_\mathrm{X}$, whereas the amount of egested faeces is denoted $M_\mathrm{P}$. All amounts of substances are accompanied by material flows such that
\begin{equation}
\dot{J}_\mathrm{*}=\dd{M_\mathrm{*}}{t},
\end{equation}
where $*$ is a wildcard symbol for either food (X), reserve (E), structure (V), or faeces (P).

Limiting the discussion to heterotrophic aerobes, the four organic-material flows are accompanied by oxygen inflow and carbon dioxide, excess water, and nitrogenous waste (e.g., ammonia, uric acid, or urea) outflows. Oxygen is used as a high-energy inorganic electron acceptor, whereas carbon dioxide, water, and nitrogenous waste are the byproducts of metabolic transformations.

Three metabolic transformations suffice to account for the accumulation of reserve, growth of structure, and organismal energy needs. `Assimilation' (A) is the conversion of food into reserve with faeces being egested in the process. `Growth' (G) is the conversion of reserve into structure. `Dissipation' (D) is the generation of metabolic energy from reserve. These ideas can formally be summarised in terms of macrochemical reactions:\\
\begin{minipage}{\textwidth}
\vspace{3mm}
\centering
\begin{tabular}{@{}l l c l@{}}
(A) & $y_\mathrm{XE}$\,CH$_{n_\mathrm{HX}}$O$_{n_\mathrm{OX}}$N$_{n_\mathrm{NX}}$ + $c_\mathrm{OA}$\,O$_2$ & $\rightarrow$ & CH$_{n_\mathrm{HE}}$O$_{n_\mathrm{OE}}$N$_{n_\mathrm{NE}}$\\
 &  &  & + $c_\mathrm{CA}$\,CO$_2$ + $c_\mathrm{HA}$\,H$_2$O + $c_\mathrm{NA}$\,NH$_3$\\
 &  &  & + $y_\mathrm{PE}$\,CH$_{n_\mathrm{HP}}$O$_{n_\mathrm{OP}}$N$_{n_\mathrm{NP}}$;\\
(G) & CH$_{n_\mathrm{HE}}$O$_{n_\mathrm{OE}}$N$_{n_\mathrm{NE}}$ + $c_\mathrm{OG}$\,O$_2$ & $\rightarrow$ & $y_\mathrm{VE}$\,CH$_{n_\mathrm{HV}}$O$_{n_\mathrm{OV}}$N$_{n_\mathrm{NV}}$\\
 &  &  & + $c_\mathrm{CG}$\,CO$_2$ + $c_\mathrm{HG}$\,H$_2$O + $c_\mathrm{NG}$\,NH$_3$;\\
(D) & CH$_{n_\mathrm{HE}}$O$_{n_\mathrm{OE}}$N$_{n_\mathrm{NE}}$ + $c_\mathrm{OD}$\,O$_2$ & $\rightarrow$ & $c_\mathrm{CD}$\,CO$_2$ + $c_\mathrm{HD}$\,H$_2$O + $c_\mathrm{ND}$\,NH$_3$.
\end{tabular}
\vspace{3mm}
\end{minipage}
Here, the stoichiometric coefficients $y_\mathrm{*E}$ are called yields. They characterise the organism's metabolism and cannot be fixed by mass-balancing the macrochemical reactions; yields are the metabolic degrees of freedom. The labels $y_\mathrm{*E}$ contain one wildcard symbol, substituting for X, V, or P.

In contrast to yields, the stoichiometric coefficients $c_\mathrm{**}$ are fixed by the mass balance:\\
\begin{minipage}{\textwidth}
\begin{minipage}[t]{0.40\textwidth}
{\begin{align*}
c_\mathrm{CA} &= y_\mathrm{XE}-1-y_\mathrm{PE}\\
c_\mathrm{HA} &= \frac{1}{2}y_\mathrm{XE}n_\mathrm{HX}-\frac{1}{2}n_\mathrm{HE}-\frac{1}{2}y_\mathrm{PE}n_\mathrm{HP}\\
 & -\frac{3}{2}y_\mathrm{XE}n_\mathrm{NX}+\frac{3}{2}n_\mathrm{NE}+\frac{3}{2}y_\mathrm{PE}n_\mathrm{NP}\\
c_\mathrm{OA} &= y_\mathrm{XE}-1-y_\mathrm{PE}\\
 & +\frac{1}{4}y_\mathrm{XE}n_\mathrm{HX}-\frac{1}{4}n_\mathrm{HE}-\frac{1}{4}y_\mathrm{PE}n_\mathrm{HP}\\
 & -\frac{1}{2}y_\mathrm{XE}n_\mathrm{OX}+\frac{1}{2}n_\mathrm{OE}+\frac{1}{2}y_\mathrm{PE}n_\mathrm{OP}\\
 & -\frac{3}{4}y_\mathrm{XE}n_\mathrm{NX}+\frac{3}{4}n_\mathrm{NE}+\frac{3}{4}y_\mathrm{PE}n_\mathrm{NP}\\
c_\mathrm{NA} &= y_\mathrm{XE}n_\mathrm{NX}-n_\mathrm{NE}-y_\mathrm{PE}n_\mathrm{NP};
\end{align*}}
\end{minipage}
\begin{minipage}[t]{0.34\textwidth}
{\begin{align*}
c_\mathrm{CG} &= 1-y_\mathrm{VE}\\
c_\mathrm{HG} &= \frac{1}{2}n_\mathrm{HE}-\frac{1}{2}y_\mathrm{VE}n_\mathrm{HV}\\
 & -\frac{3}{2}n_\mathrm{NE}+\frac{3}{2}y_\mathrm{VE}n_\mathrm{NV}\\
c_\mathrm{OG} &= 1-y_\mathrm{VE}\\
 & +\frac{1}{4}n_\mathrm{HE}-\frac{1}{4}y_\mathrm{VE}n_\mathrm{HV}\\
 & -\frac{1}{2}n_\mathrm{OE}+\frac{1}{2}y_\mathrm{VE}n_\mathrm{OV}\\
 & -\frac{3}{4}n_\mathrm{NE}+\frac{3}{4}y_\mathrm{VE}n_\mathrm{NV}\\
c_\mathrm{NG} &= n_\mathrm{NE}-y_\mathrm{VE}n_\mathrm{NV};
\end{align*}}
\end{minipage}
\begin{minipage}[t]{0.24\textwidth}
{\begin{align*}
c_\mathrm{CD} &= 1\\
c_\mathrm{HD} &= \frac{1}{2}n_\mathrm{HE}-\frac{3}{2}n_\mathrm{NE}\\
c_\mathrm{OD} &= 1+\frac{1}{4}n_\mathrm{HE}\\
 & -\frac{1}{2}n_\mathrm{OE}-\frac{3}{4}n_\mathrm{NE}\\
c_\mathrm{ND} &= n_\mathrm{NE}.
\end{align*}}
\end{minipage}
\vspace{3mm}
\end{minipage}
The labels $c_\mathrm{**}$ contain two wildcard symbols. The first one stands either for carbon (C), hydrogen (H), oxygen (O), or nitrogen (N), whereas the second one substitutes for A, G, or D. The chemical indices $n_\mathrm{**}$, which specify the macromolecular composition of food, reserve, structure, and faeces, also come with two wildcard symbols. The first one substitutes for H, O, or N (but not C because working in C-moles implies that $n_\mathrm{C*}=1$), whereas the second one substitutes for X, E, V, or P. The four elements C, H, O, and N typically constitute 99\,\% of organismal mass balance, but adding other elements is also possible when specific applications call for it. 

The mass balance can be rewritten in a more compact way using matrix notation. This becomes apparent by arranging the organic macromolecules (X, E, V, and P) and the metabolites (CO$_2$, H$_2$O, O$_2$, and NH$_3$) all on the the right-hand side of the macrochemical reactions. If we now define
\begin{align}
\mathbf{Y}=
\begin{blockarray}{cccl}
\mathrm{A} & \mathrm{G} & \mathrm{D}\\
\begin{block}{(ccc)l}
-y_\mathrm{XE} &              0 &   0 & \mathrm{X}\\
             1 &             -1 &  -1 & \mathrm{E}\\
             0 &  y_\mathrm{VE} &   0 & \mathrm{V}\\
 y_\mathrm{PE} &              0 &   0 & \mathrm{P}\\
\end{block}
\end{blockarray}
& \hspace{0mm}
& \mathbf{C}=
\begin{blockarray}{cccl}
\mathrm{A} & \mathrm{G} & \mathrm{D}\\
\begin{block}{(ccc)l}
 c_\mathrm{CA} &  c_\mathrm{CG} &  c_\mathrm{CD} & \mathrm{CO}_2\\
 c_\mathrm{HA} &  c_\mathrm{HG} &  c_\mathrm{HD} & \mathrm{H}_2\mathrm{O}\\
-c_\mathrm{OA} & -c_\mathrm{OG} & -c_\mathrm{OD} & \mathrm{O}_2\\
 c_\mathrm{NA} &  c_\mathrm{NG} &  c_\mathrm{ND} & \mathrm{NH}_3\\
\end{block}
\end{blockarray}
\end{align}
and
\begin{align}
\mathbf{N}_\mathrm{org}=
\begin{blockarray}{ccccl}
\mathrm{X} & \mathrm{E} & \mathrm{V} & \mathrm{P}\\
\begin{block}{(cccc)l}
            1 &             1 &             1 &             1 & \mathrm{C}\\
n_\mathrm{HX} & n_\mathrm{HE} & n_\mathrm{HV} & n_\mathrm{HP} & \mathrm{H}\\
n_\mathrm{OX} & n_\mathrm{OE} & n_\mathrm{OV} & n_\mathrm{OP} & \mathrm{O}\\
n_\mathrm{NX} & n_\mathrm{OE} & n_\mathrm{NV} & n_\mathrm{NP} & \mathrm{N}\\
\end{block}
\end{blockarray}
& \hspace{0mm}
& \mathbf{N}_\mathrm{met}=
\begin{blockarray}{ccccl}
\mathrm{CO}_2 & \mathrm{H}_2\mathrm{O} & \mathrm{O}_2 & \mathrm{NH}_3\\
\begin{block}{(cccc)l}
1 & 0 & 0 & 0 & \mathrm{C}\\
0 & 2 & 0 & 3 & \mathrm{H}\\
2 & 1 & 2 & 0 & \mathrm{O}\\
0 & 0 & 0 & 1 & \mathrm{N}\\
\end{block}
\end{blockarray},
\end{align}
then the mass balance becomes
\begin{align}
\mathbf{0}=\mathbf{N}_\mathrm{org}\mathbf{Y}+\mathbf{N}_\mathrm{met}\mathbf{C}
& \hspace{0mm}
& \text{or} %\Leftrightarrow
& \hspace{0mm}
& \mathbf{C}=-\mathbf{N}_\mathrm{met}^{-1}\mathbf{N}_\mathrm{org}\mathbf{Y}.
\label{eq:massbalance}
\end{align}
The minus signs in matrices $\mathbf{Y}$ and $\mathbf{C}$ arise from transferring the organic macromolecules and oxygen to the right-hand side of the macrochemical reactions. The matrix $\mathbf{N}_\mathrm{met}$ is invertible.

Eq.~\eqref{eq:massbalance} explicitly shows that metabolism is characterised by three free parameters, that is, the yields $y_\mathrm{*E}$. This is also reflected in the fact that all material flows can be expressed as the linear combinations of three basic flows, the inflow into reserve and the outflows from reserve. The assimilation macrochemical reaction shows that in time it takes to assimilate 1 C-mole of reserve, the organism ingests $y_\mathrm{XE}$ C-moles of food and egests $y_\mathrm{PE}$ C-moles of faeces. Accordingly, if the rate of reserve assimilation is $\dot{J}_\mathrm{EA}$, the ingestion rate is $\dot{J}_\mathrm{X}=y_\mathrm{XE}\dot{J}_\mathrm{EA}$ and the egestion rate is $\dot{J}_\mathrm{P}=y_\mathrm{PE}\dot{J}_\mathrm{EA}$. The growth macrochemical reaction further implies that, if $\dot{J}_\mathrm{EG}$ is the rate of reserve utilisation for growth, then the structural growth rate is $\dot{J}_\mathrm{V}=y_\mathrm{VE}\dot{J}_\mathrm{EG}$. Besides growth, reserve is also utilised for generating metabolic energy at the rate $\dot{J}_\mathrm{ED}$, meaning that reserve is built at the net rate $\dot{J}_\mathrm{E}=\dot{J}_\mathrm{EA}-\dot{J}_\mathrm{EG}-\dot{J}_\mathrm{ED}$. Summarising these considerations using matrix notation:
\begin{align}
\dot{\mathbf{J}}_\mathrm{org}=\mathbf{Y}\dot{\mathbf{J}}_\mathrm{E},
&\hspace{0mm}
&\text{where}
&\hspace{0mm}
&\dot{\mathbf{J}}_\mathrm{org}=
\begin{pmatrix}
-\dot{J}_\mathrm{X}\\
 \dot{J}_\mathrm{E}\\
 \dot{J}_\mathrm{V}\\
 \dot{J}_\mathrm{P}
\end{pmatrix}
&\hspace{0mm}
&\text{and}
&\hspace{0mm}
&\dot{\mathbf{J}}_\mathrm{E}=
\begin{pmatrix}
\dot{J}_\mathrm{EA}\\
\dot{J}_\mathrm{EG}\\
\dot{J}_\mathrm{ED}
\end{pmatrix}.
\label{eq:3centralA}
\end{align}
Analogous reasoning applied to the four metabolites leads to
\begin{align}
\dot{\mathbf{J}}_\mathrm{met}=\mathbf{C}\dot{\mathbf{J}}_\mathrm{E}=-\mathbf{N}_\mathrm{met}^{-1}\mathbf{N}_\mathrm{org}\mathbf{Y}\dot{\mathbf{J}}_\mathrm{E},
&\hspace{0mm}
&\text{where}
&\hspace{0mm}
&\dot{\mathbf{J}}_\mathrm{met}=
\begin{pmatrix}
 \dot{J}_\mathrm{C}\\
 \dot{J}_\mathrm{H}\\
-\dot{J}_\mathrm{O}\\
 \dot{J}_\mathrm{N}\\
\end{pmatrix}.
\label{eq:3centralB}
\end{align}
The material flows in the $\dot{\mathbf{J}}_\mathrm{met}$ matrix should respectively be interpreted as the rate of carbon-dioxide outflow, the rate of metabolic water production, the rate of oxygen inflow, and the rate of metabolic nitrogenous-waste production.

\subsection{Heat balance}

To arrive at the organismal heat balance, it is necessary to establish how the amounts and flows of substances relate to energy and the flows thereof. This is achieved by starting from the fundamental thermodynamic relation, which unlike the laws in Box~1 of the main text, does not connect changes in the control volume strictly to flows across the control surface, but rather is a statement about the state of the control volume itself. Given a constant temperature $T$ and pressure $p$ (implying that only near-equilibrium, quasi-static processes take place), changes in internal energy $\mathrm{d}U$ are
\begin{equation}
\mathrm{d}U = T\mathrm{d}S - p\mathrm{d}V + \overline{\bm{\upmu}}^\mathrm{T} \mathrm{d}\mathbf{M},
\label{eq:simple1stlaw}
\end{equation}
where $\mathrm{d}S$ is entropy change, and the terms $-p \mathrm{d}V$ and $\overline{\bm{\upmu}}^\mathrm{T} \mathrm{d}\mathbf{M}$ respectively represent the expansion work and substance conversions ($V$ is the size of the control volume and $\overline{\bm{\upmu}}$ is a vector of chemical potentials; see below). Rearranging Eq.~\eqref{eq:simple1stlaw} leads to
\begin{align}
\mathrm{d}\lr{U+pV-TS} &= \overline{\bm{\upmu}}^\mathrm{T} \mathrm{d}\mathbf{M}
& \hspace{0mm}
& \text{or} %\Leftrightarrow
& \hspace{0mm}
\mathrm{d}G &= \overline{\bm{\upmu}}^\mathrm{T} \mathrm{d}\mathbf{M},
\end{align}
where $G=U+pV-TS$ is the Gibbs free energy, a thermodynamic state function whose changes have a desirable property of being proportional to changes in the amounts of substances.

In the context of DEB theory, we keep track of four Gibbs free energies:
\begin{subequations}
\begin{align}
E_\mathrm{X} &=\overline{\mu}_\mathrm{X} M_\mathrm{X} &\hspace{0mm} &\text{food};\\
E            &=\overline{\mu}_\mathrm{E} M_\mathrm{E} &\hspace{0mm} &\text{reserve};\\
E_\mathrm{V} &=\overline{\mu}_\mathrm{V} M_\mathrm{V} &\hspace{0mm} &\text{structure};\\
E_\mathrm{P} &=\overline{\mu}_\mathrm{P} M_\mathrm{P} &\hspace{0mm} &\text{faeces},
\end{align}
\end{subequations}
where $\overline{\mu}_\mathrm{*}$ are the chemical potentials, each converting an amount of substance to the corresponding energy. We furthermore keep track of three energy flows:
\begin{subequations}
\begin{align}
\dot{p}_\mathrm{A} &=\overline{\mu}_\mathrm{E}\dot{J}_\mathrm{EA} &\hspace{0mm} &\text{assimilation;}\\
\dot{p}_\mathrm{G} &=\overline{\mu}_\mathrm{E}\dot{J}_\mathrm{EG} &\hspace{0mm} &\text{growth;}\\
\dot{p}_\mathrm{D} &=\overline{\mu}_\mathrm{E}\dot{J}_\mathrm{ED} &\hspace{0mm} &\text{dissipation}.
\end{align}
\end{subequations}
Restricting attention to assimilation, growth, and dissipation flows is possible thanks to the aforementioned central role of the inflow into and the outflows from reserve expressed by Eqs.~\eqref{eq:3centralA} and \eqref{eq:3centralB}. We can now extend the matrix notation to write
\begin{align}
\dot{\mathbf{p}}=\overline{\mu}_\mathrm{E}\dot{\mathbf{J}}_\mathrm{E}
&\hspace{0mm}
&\text{or}
&\hspace{0mm}
&\dot{\mathbf{J}}_\mathrm{E}=\frac{1}{\overline{\mu}_\mathrm{E}}\dot{\mathbf{p}},
&\hspace{0mm}
&\text{where}
&\hspace{0mm}
&\dot{\mathbf{p}}=
\begin{pmatrix}
\dot{p}_\mathrm{A}\\
\dot{p}_\mathrm{G}\\
\dot{p}_\mathrm{D}
\end{pmatrix}.
\end{align}

Heat balance is obtained by applying Eq.~(1) of the main text in the context of DEB theory, that is, to the control volume comprising reserve and structure. Tracking all inflows into and outflows from the control volume yields
\begin{align}
\overline{\mathbf{h}}^\mathrm{T}\dd{\mathbf{M}}{t} &\mapsto \sum_\mathrm{in}{\overline{h}_\mathrm{*}\dd{M_\mathrm{*}}{t}} - \sum_\mathrm{out}{\overline{h}_\mathrm{*}\dd{M_\mathrm{*}}{t}}\nonumber\\
&= \overline{h}_\mathrm{X}\dot{J}_\mathrm{X} + \overline{h}_\mathrm{O}\dot{J}_\mathrm{O} - \overline{h}_\mathrm{P}\dot{J}_\mathrm{P} - \overline{h}_\mathrm{C}\dot{J}_\mathrm{C} - 
\overline{h}_\mathrm{H}\dot{J}_\mathrm{H} - \overline{h}_\mathrm{N}\dot{J}_\mathrm{N}.
\end{align}
Inside the control volume, the buildup of reserve and structure (at constant pressure) causes temporal changes in enthalpy:
\begin{equation}
\dd{H}{t} \mapsto \overline{h}_\mathrm{E}\dd{M_\mathrm{E}}{t} + \overline{h}_\mathrm{V}\dd{M_\mathrm{V}}{t} = \overline{h}_\mathrm{E}\dot{J}_\mathrm{E} + \overline{h}_\mathrm{V}\dot{J}_\mathrm{V}.
\end{equation}
Inserting the last two expressions into Eq.~(1) while accounting for the definition enthalpy, $H=U+pV$, we get
\begin{equation}
-\dot{Q} = \dot{W}' + \overline{h}_\mathrm{X}\dot{J}_\mathrm{X} - \overline{h}_\mathrm{E}\dot{J}_\mathrm{E} - \overline{h}_\mathrm{V}\dot{J}_\mathrm{V} - \overline{h}_\mathrm{P}\dot{J}_\mathrm{P} - \overline{h}_\mathrm{C}\dot{J}_\mathrm{C} - 
\overline{h}_\mathrm{H}\dot{J}_\mathrm{H} + \overline{h}_\mathrm{O}\dot{J}_\mathrm{O} -
\overline{h}_\mathrm{N}\dot{J}_\mathrm{N},
\label{eq:heatgen}
\end{equation}
where $\dot{Q}$ and $\dot{W}'$ respectively denote the amounts of heat and non-expansionary work per unit of time. It is again convenient to resort to matrix notation, to which end we introduce vectors $\mathbf{h}^\mathrm{T}_\mathrm{org} = \lr{\overline{h}_\mathrm{X}, \overline{h}_\mathrm{E}, \overline{h}_\mathrm{V}, \overline{h}_\mathrm{P}}$ and $\mathbf{h}^\mathrm{T}_\mathrm{met} = \lr{\overline{h}_\mathrm{C}, \overline{h}_\mathrm{H}, \overline{h}_\mathrm{O}, \overline{h}_\mathrm{N}}$, thus obtaining
\begin{align}
-\dot{Q} &= \dot{W}' - \mathbf{h}^\mathrm{T}_\mathrm{org} \dot{\mathbf{J}}_\mathrm{org} - \mathbf{h}^\mathrm{T}_\mathrm{met} \dot{\mathbf{J}}_\mathrm{met}\nonumber\\
&= \dot{W}' + \lr{\mathbf{h}^\mathrm{T}_\mathrm{met}\mathbf{N}_\mathrm{met}^{-1}\mathbf{N}_\mathrm{org} - \mathbf{h}^\mathrm{T}_\mathrm{org}} \mathbf{Y}\dot{\mathbf{J}}_\mathrm{E}\nonumber\\
&= \dot{W}' + \frac{1}{\overline{\mu}_\mathrm{E}} \lr{\mathbf{h}^\mathrm{T}_\mathrm{met}\mathbf{N}_\mathrm{met}^{-1}\mathbf{N}_\mathrm{org} - \mathbf{h}^\mathrm{T}_\mathrm{org}}\mathbf{Y}\dot{\mathbf{p}}.
\end{align}

\subsection{Entropy balance}

Arriving at the organismal entropy balance involves similar reasoning as for the heat balance, but the starting point is Eq.~(3) of the main text. Applying this equation in the context of DEB theory, that is, to the control volume comprising reserve and structure, we have
\begin{subequations}
\begin{align}
\overline{\bm{\upmu}}^\mathrm{T}\dd{\mathbf{M}}{t} &\mapsto \sum_\mathrm{in}{\overline{\mu}_\mathrm{*}\dd{M_\mathrm{*}}{t}} - \sum_\mathrm{out}{\overline{\mu}_\mathrm{*}\dd{M_\mathrm{*}}{t}}\nonumber\\ &= \overline{\mu}_\mathrm{X}\dot{J}_\mathrm{X} + \overline{\mu}_\mathrm{O}\dot{J}_\mathrm{O} - \overline{\mu}_\mathrm{P}\dot{J}_\mathrm{P} - \overline{\mu}_\mathrm{C}\dot{J}_\mathrm{C} - 
\overline{\mu}_\mathrm{H}\dot{J}_\mathrm{H} - \overline{\mu}_\mathrm{N}\dot{J}_\mathrm{N};\\
\dd{G}{t} &\mapsto \overline{\mu}_\mathrm{E}\dd{M_\mathrm{E}}{t} + \overline{\mu}_\mathrm{V}\dd{M_\mathrm{V}}{t} = \overline{\mu}_\mathrm{E}\dot{J}_\mathrm{E} + \overline{\mu}_\mathrm{V}\dot{J}_\mathrm{V}.
\end{align}
\end{subequations}
Inserting these expressions into Eq.~(3) gives
\begin{equation}
T\dot{\sigma} = \dot{W}' + \overline{\mu}_\mathrm{X}\dot{J}_\mathrm{X} - \overline{\mu}_\mathrm{E}\dot{J}_\mathrm{E} - \overline{\mu}_\mathrm{V}\dot{J}_\mathrm{V} - \overline{\mu}_\mathrm{P}\dot{J}_\mathrm{P} - \overline{\mu}_\mathrm{C}\dot{J}_\mathrm{C} - 
\overline{\mu}_\mathrm{H}\dot{J}_\mathrm{H} + \overline{\mu}_\mathrm{O}\dot{J}_\mathrm{O} -
\overline{\mu}_\mathrm{N}\dot{J}_\mathrm{N},
\label{eq:enprod}
\end{equation}
where $\dot{\sigma}$ denotes the amount of entropy production per unit of time. Using matrix notation with definitions $\mathbf{\bm{\upmu}}^\mathrm{T}_\mathrm{org} = \lr{\overline{\mu}_\mathrm{X}, \overline{\mu}_\mathrm{E}, \overline{\mu}_\mathrm{V}, \overline{\mu}_\mathrm{P}}$ and $\mathbf{\bm{\upmu}}^\mathrm{T}_\mathrm{met} = \lr{\overline{\mu}_\mathrm{C}, \overline{\mu}_\mathrm{H}, \overline{\mu}_\mathrm{O}, \overline{\mu}_\mathrm{N}}$, we can write
\begin{align}
T\dot{\sigma} &= \dot{W}' -\mathbf{\bm{\upmu}}^\mathrm{T}_\mathrm{org} \dot{\mathbf{J}}_\mathrm{org} - \mathbf{\bm{\upmu}}^\mathrm{T}_\mathrm{met} \dot{\mathbf{J}}_\mathrm{met}\nonumber\\
&= \dot{W}' + \lr{\mathbf{\bm{\upmu}}^\mathrm{T}_\mathrm{met}\mathbf{N}_\mathrm{met}^{-1}\mathbf{N}_\mathrm{org} - \mathbf{\bm{\upmu}}^\mathrm{T}_\mathrm{org}} \mathbf{Y}\dot{\mathbf{J}}_\mathrm{E}\nonumber\\
&= \dot{W}' + \frac{1}{\overline{\mu}_\mathrm{E}} \lr{\mathbf{\bm{\upmu}}^\mathrm{T}_\mathrm{met}\mathbf{N}_\mathrm{met}^{-1}\mathbf{N}_\mathrm{org} - \mathbf{\bm{\upmu}}^\mathrm{T}_\mathrm{org}}\mathbf{Y}\dot{\mathbf{p}}.
\end{align}

Of note is that $\dot{W}'<0$ because the organism performs work on the environment at the expense of internal energy. If the organism is very active, $T\dot{\sigma}$ seemingly turns negative, which is physically impossible. Energy flows must therefore compensate for intense activity; the organism needs to dissipate more, and eventually eat more. Assuming that the excess dissipation $\Delta\dot{p}_\mathrm{D}$ is due to the physical work $\dot{W}'$, a typical relationship between these two quantities will be $\dot{W}'=-\eta \Delta\dot{p}_\mathrm{D}$, where \smash{$0<\eta<1 - \frac{\overline{\mu}_\mathrm{C}}{\overline{\mu}_\mathrm{E}}c_\mathrm{CD} - \frac{\overline{\mu}_\mathrm{H}}{\overline{\mu}_\mathrm{E}}c_\mathrm{HD} + \frac{\overline{\mu}_\mathrm{O}}{\overline{\mu}_\mathrm{E}}c_\mathrm{OD} - \frac{\overline{\mu}_\mathrm{N}}{\overline{\mu}_\mathrm{E}}c_\mathrm{ND}$} is muscle efficiency.

\subsection{Molar thermochemical quantities}

Having derived the expressions for heat generation and entropy production from first principles, it is necessary to address the question as to what the typical chemical potentials and molar enthalpies are. To find the appropriate values for these quantities, we first show that the right-hand sides of Eqs.~\eqref{eq:heatgen} and \eqref{eq:enprod} are what thermochemists would respectively recognise as (negative) enthalpy and Gibbs free energy of a reaction. The reaction in question is the aggregate of the three metabolic transformations: assimilation, growth, and dissipation.

Thermochemists define enthalpy of a reaction as \smash{$\Delta_\mathrm{rxn}H = {\sum c_\mathrm{prod}\overline{h}_\mathrm{prod}} - {\sum c_\mathrm{reac}\overline{h}_\mathrm{reac}}$}, where $c_\mathrm{prod}$ and $c_\mathrm{reac}$ are the stoichiometric coefficients accompanying products and reactants, respectively. The quantities $\overline{h}_\mathrm{prod}$ and $\overline{h}_\mathrm{reac}$ are the enthalpies of products and reactants, respectively. An analogous definition holds for the Gibbs free energy of a reaction, \smash{$\Delta_\mathrm{rxn}G = {\sum c_\mathrm{prod}\overline{\mu}_\mathrm{prod}} - {\sum c_\mathrm{reac}\overline{\mu}_\mathrm{reac}}$}. 

Focusing on macrochemical reactions inside the control volume, we have
\begin{align}
\Delta_\mathrm{rnx}^\mathrm{c.v.}G = -\frac{T\dot{\sigma}}{\dot{J}_\mathrm{EA}}\nonumber &= -\overline{\mu}_\mathrm{X}\frac{\dot{J}_\mathrm{X}}{\dot{J}_\mathrm{EA}} + \overline{\mu}_\mathrm{E}\frac{\dot{J}_\mathrm{E}}{\dot{J}_\mathrm{EA}} + \overline{\mu}_\mathrm{V}\frac{\dot{J}_\mathrm{V}}{\dot{J}_\mathrm{EA}} + \overline{\mu}_\mathrm{P}\frac{\dot{J}_\mathrm{P}}{\dot{J}_\mathrm{EA}}\nonumber\\
&+ \overline{\mu}_\mathrm{C}\frac{\dot{J}_\mathrm{C}}{\dot{J}_\mathrm{EA}} + 
\overline{\mu}_\mathrm{H}\frac{\dot{J}_\mathrm{H}}{\dot{J}_\mathrm{EA}} - \overline{\mu}_\mathrm{O}\frac{\dot{J}_\mathrm{O}}{\dot{J}_\mathrm{EA}} +
\overline{\mu}_\mathrm{N}\frac{\dot{J}_\mathrm{N}}{\dot{J}_\mathrm{EA}}\nonumber\\
&= \overline{\mu}_\mathrm{E}\lr{1-\frac{\dot{J}_\mathrm{EG}}{\dot{J}_\mathrm{EA}}-\frac{\dot{J}_\mathrm{ED}}{\dot{J}_\mathrm{EA}}}
+ \overline{\mu}_\mathrm{V}y_\mathrm{VE}\frac{\dot{J}_\mathrm{EG}}{\dot{J}_\mathrm{EA}}
+ \overline{\mu}_\mathrm{P}y_\mathrm{PE}\nonumber\\
&+ \overline{\mu}_\mathrm{C}\lr{c_\mathrm{CA}+c_\mathrm{CG}\frac{\dot{J}_\mathrm{EG}}{\dot{J}_\mathrm{EA}}+c_\mathrm{CD}\frac{\dot{J}_\mathrm{ED}}{\dot{J}_\mathrm{EA}}}\nonumber\\
&+ \overline{\mu}_\mathrm{H}\lr{c_\mathrm{HA}+c_\mathrm{HG}\frac{\dot{J}_\mathrm{EG}}{\dot{J}_\mathrm{EA}}+c_\mathrm{HD}\frac{\dot{J}_\mathrm{ED}}{\dot{J}_\mathrm{EA}}}\nonumber\\
&+ \overline{\mu}_\mathrm{N}\lr{c_\mathrm{NA}+c_\mathrm{NG}\frac{\dot{J}_\mathrm{EG}}{\dot{J}_\mathrm{EA}}+c_\mathrm{ND}\frac{\dot{J}_\mathrm{ED}}{\dot{J}_\mathrm{EA}}}\nonumber\\
&- \overline{\mu}_\mathrm{X}y_\mathrm{XE}
- \overline{\mu}_\mathrm{O}\lr{c_\mathrm{OA}+c_\mathrm{OG}\frac{\dot{J}_\mathrm{EG}}{\dot{J}_\mathrm{EA}}+c_\mathrm{OD}\frac{\dot{J}_\mathrm{ED}}{\dot{J}_\mathrm{EA}}}.
\end{align}
Here, reserve (E), structure (V), faeces (P), carbon dioxide (C), water (H), and nitrogenous waste (N) are products, whereas food (X) and oxygen (O) are reactants. Each chemical potential is accompanied by a stoichiometric coefficient that arises from aggregating the three macrochemical reactions. The ratios \smash{$\frac{\dot{J}_\mathrm{EG}}{\dot{J}_\mathrm{EA}}$} and \smash{$\frac{\dot{J}_\mathrm{ED}}{\dot{J}_\mathrm{EA}}$} respectively adjust for the rates at which growth and dissipation occur relative to assimilation. We can similarly establish that \smash{$\Delta_\mathrm{rnx}^\mathrm{c.v.}H = \frac{\dot{Q}}{\dot{J}_\mathrm{EA}}$}. A practical implication of these considerations is that we should simply follow thermochemical principles to find the appropriate values for molar thermodynamic quantities.

Molar thermodynamic quantities cannot be measured in absolute terms, but rather relative to a reference state. Thermochemistry most commonly uses the standard state as reference. Skipping much of the details, the standard state is that of a pure substance subjected to the pressure of 10$^5$\,Pa. When referring to heterotrophic aerobes, however, it is more convenient to use the products of combustion as reference, specifically CO$_2$, H$_2$O, and N$_2$. This is because living aerobic cells obtain energy through the breakdown of organic substances to CO$_2$ and H$_2$O. We also have that $\overline{\mu} = \Delta_\mathrm{c}G_\mathrm{m}$ and $\overline{h} = \Delta_\mathrm{c}H_\mathrm{m}$, where the quantities $\Delta_\mathrm{c}G_\mathrm{m}$ and $\Delta_\mathrm{c}H_\mathrm{m}$, respectively called molar Gibbs free energy and molar enthalpy of combustion, are readily measurable via calorimetry. They are also calculable from standard thermodynamic tables by reversing the combustion reaction for 1 mole or C-mole of a chosen substance, and then calculating the corresponding Gibbs free energy or enthalpy of reaction. This is exemplified in the case of glucose:\\
\begin{minipage}{\textwidth}
\vspace{3mm}
\centering
\begin{tabular}{@{} l c l @{}}
CO$_2$ + H$_2$O & $\rightarrow$ & CH$_2$O + $\frac{3}{2}$O$_2$,
\end{tabular}
\vspace{3mm}
\end{minipage}
implying that
\begin{align*}
\Delta_\mathrm{c}G_\mathrm{m}&=\lrs{1\lr{-152.6}+\frac{3}{2}\cdot 0 - 1\lr{-394.4} - 1\lr{-237.1}}\,\text{J\,C-mol}^{-1}=478.9\,\text{J\,C-mol}^{-1};\\
\Delta_\mathrm{c}H_\mathrm{m}&=\lrs{1\lr{-212.217}+\frac{3}{2}\cdot 0 - 1\lr{-393.51} - 1\lr{-285.83}}\,\text{J\,C-mol}^{-1}=467.123\,\text{J\,C-mol}^{-1}.
\end{align*}

\begin{figure}[t]
\includegraphics[scale=1]{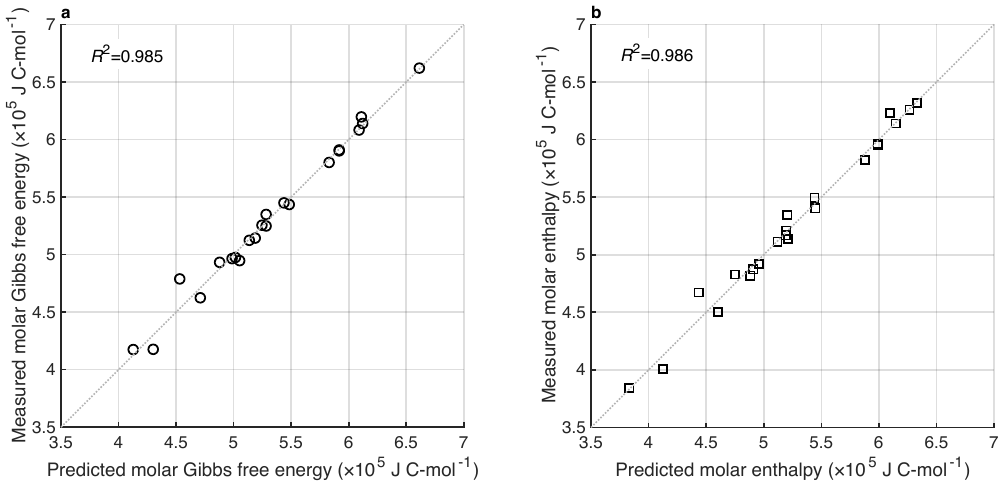}
\centering
\caption{Estimating molar thermochemical quantities from the chemical composition of substances. (\textbf{a}) Measured vs. estimated plot for molar Gibbs free energy (i.e., chemical potential). (\textbf{b}) Measured vs. estimated plot for molar enthalpy. The measured values are from Table~\ref{tab:combustion}, whereas the estimated values are from a multiple linear regression model that takes the chemical indices $n_\mathrm{**}$ as inputs (see text for details).}
\label{fig:regress}
\end{figure}

A list of molar Gibbs free energies and molar enthalpies of combustion is given in Table~\ref{tab:combustion}. In most instances $\Delta_\mathrm{c}G_\mathrm{m} \approx \Delta_\mathrm{c}H_\mathrm{m}$, indicating that for heterotrophic aerobes $-\dot{Q} \approx T\dot{\sigma}$, as emphasised in the main text.

Of note is that the choice of the reference state in thermochemistry is arbitrary; for example, carbon dioxide and elementary nitrogen gas in our definition may be replaced by bicarbonate and ammonium ions, respectively, which somewhat changes the values listed in Table~\ref{tab:combustion}. Caution must be exercised with respect to the reference state when extracting the values of molar Gibbs free energies and molar enthalpies of combustion from literature.

In applications, it is possible to arrive at a quick estimate of molar thermochemical properties based solely on the chemical composition of an organic substance (Figure~\ref{fig:regress}). A multiple linear regression model that takes the chemical indices $n_\mathrm{**}$ as inputs is sufficient for this task. The model takes the form:
\begin{subequations}
\begin{align}
\Delta_\mathrm{c}G_\mathrm{m} &= \alpha_\mathrm{C} + \alpha_\mathrm{H} n_\mathrm{H*} + \alpha_\mathrm{O} n_\mathrm{O*} + \alpha_\mathrm{N} n_\mathrm{N*};\\
\Delta_\mathrm{c}H_\mathrm{m} &= \beta_\mathrm{C} + \beta_\mathrm{H} n_\mathrm{H*} + \beta_\mathrm{O} n_\mathrm{O*} + \beta_\mathrm{N} n_\mathrm{N*},
\end{align}
\end{subequations}
where the regression coefficients (and the corresponding 95\,\% confidence intervals) are\\
\begin{minipage}{\textwidth}
\vspace{3mm}
\centering
\begin{tabular}{@{}l l@{}}
$\alpha_\mathrm{C} = 4.384~\lr{4.249, 4.520}$ & $\beta_\mathrm{C} = 4.328~\lr{4.182, 4.474}$;\\
$\alpha_\mathrm{H} = 0.982~\lr{0.904, 1.060}$ & $\beta_\mathrm{H} = 1.099~\lr{1.016, 1.183}$;\\
$\alpha_\mathrm{O} = -1.818~\lr{-1.961, -1.675}$ & $\beta_\mathrm{O} = -2.091~\lr{-2.245, -1.938}$;\\
$\alpha_\mathrm{N} = 0.059~\lr{-0.057, 0.175}$ & $\beta_\mathrm{N} = -0.151~\lr{-0.276, -0.026}$.
\end{tabular}
\vspace{3mm}
\end{minipage}
These regression coefficients resemble the valences of the elements C, H, and O (equal to 4, 1, and 2, respectively). The same is not the case for nitrogen N whose valence is typically, including in all amino acids, 3.

\begin{table}[p]
\centering
\caption{List of molar Gibbs free energies and molar enthalpies of combustion for select inorganic (carbon dioxide, water, and ammonia) and organic (urea and uric acid, glucose, palmitic acid, and 20 amino acids) substances.}
\vspace{3mm}
\begin{tabular}{l l c c}
Compound & Composition & $\Delta_\mathrm{c}G_\mathrm{m}$ (kJ\,mol$^{-1}$) & $\Delta_\mathrm{c}H_\mathrm{m}$ (kJ\,mol$^{-1}$)\\
\hline
Carbon dioxide  & CO$_2$(g)                                 & 0       & 0       \\
Water           & H$_2$O(g)                                 & 8.5     & 44.004  \\
                & H$_2$O(l)                                 & 0       & 0       \\
Ammonia         & NH$_3$(g)                                 & 339.25  & 382.805 \\
                & NH$_4^+$(aq)                              & 392.1   & 438.923 \\
Urea            & CH$_4$ON$_2$(aq)(s)                       & 662.2   & 631.89  \\
Uric acid       & CH$_{0.8}$O$_{0.6}$N$_{0.8}$(aq)(s)       & 417.48  & 384.238 \\
Glucose         & CH$_2$O(aq)(s)                            & 478.9   & 467.123 \\
Palmitic acid   & CH$_2$O$_{0.125}$(aq)(l)                  & 614.025 & 625.975 \\
Serine          & CH$_{2.333}$ON$_{0.333}$(aq)(s)           & 493.333 & 482.757 \\
Leucine         & CH$_{2.167}$O$_{0.333}$N$_{0.167}$(aq)(l) & 590.2   & 594.85  \\
Alanine         & CH$_{2.333}$O$_{0.667}$N$_{0.333}$(aq)(s) & 543.6   & 540.333 \\
Glycine         & CH$_{2.5}$ON$_{0.5}$(aq)(s)               & 494.8   & 487.25  \\
Lysine          & CH$_{2.333}$O$_{0.333}$N$_{0.333}$(aq)(s) & 608.333 & 613.883 \\
Valine          & CH$_{2.2}$O$_{0.4}$N$_{0.2}$(aq)(s)       & 580.16  & 582.22  \\
Threonine       & CH$_{2.25}$O$_{0.75}$N$_{0.25}$(aq)(s)    & 525.575 & 520.925 \\
Aspartic acid   & CH$_{1.75}$ON$_{0.25}$(aq)(s)             & 417.7   & 400.67  \\
Glutamic acid   & CH$_{1.8}$O$_{0.8}$N$_{0.2}$(aq)(s)       & 462.48  & 450.34  \\
Proline         & CH$_{1.8}$O$_{0.4}$N$_{0.2}$(aq)(s)       & 545.02  & 549.48  \\
Asparagine      & CH$_{2}$O$_{0.75}$N$_{0.5}$(aq)(s)        & 497.55  & 481.745 \\
Arginine        & CH$_{2.333}$O$_{0.333}$N$_{0.667}$(aq)(s) & 619.75  & 623.067 \\
Phenylalanine   & CH$_{1.222}$O$_{0.222}$N$_{0.111}$(aq)(s) & 514.422 & 516.822 \\
Isoleucine      & CH$_{2.167}$O$_{0.333}$N$_{0.167}$(aq)(s) & 591.033 & 596.233 \\
Glutamine       & CH$_2$O$_{0.6}$N$_{0.4}$(aq)(s)           & 524.9   & 513.724 \\
Cysteine        & CH$_{2.333}$O$_{0.667}$N$_{0.333}$S$_{0.333}$(aq)(s) & 677.567$^\mathrm{a}$ & 680.973$^\mathrm{a}$ \\
Tyrosine        & CH$_{1.222}$O$_{0.333}$N$_{0.111}$(aq)(s) & 496.544 & 491.822 \\
Histidine       & CH$_{1.5}$O$_{0.333}$N$_{0.5}$(aq)(s)     & 534.967 & 534.367 \\
Methionine      & CH$_{2.2}$O$_{0.4}$N$_{0.2}$S$_{0.2}$(aq)(s) & 662.42$^\mathrm{a}$ & 671.5$^\mathrm{a}$ \\
Tryptophan      & CH$_{1.091}$O$_{0.182}$N$_{0.182}$(aq)(s) & 512.455 & 511.227 \\
Avg. amino acid & CH$_{2.070}$O$_{0.595}$N$_{0.301}$S$_{0.015}${}$^\mathrm{b}$ & 538.696$^\mathrm{a}$ & 535.667$^\mathrm{a}$ \\
Avg. amino acid & CH$_{1.976}$O$_{0.492}$N$_{0.278}$S$_{0.01}${}$^\mathrm{c}$ & --- & 595.755$^\mathrm{a}$ \\
Avg. biomass    & CH$_{1.8}$O$_{0.5}$N$_{0.2}${}$^\mathrm{d}$ & 541.2$^\mathrm{d}$ & 560$^\mathrm{d}$ \\
\hline
\multicolumn{4}{l}{$^\mathrm{a}$Combustion assumed to produce SO$_3$(g).}\\
\multicolumn{4}{l}{$^\mathrm{b}$Calculated using amino acid frequencies in vertebrates}\\
\multicolumn{4}{l}{~~\url{http://www.tiem.utk.edu/~gross/bioed/webmodules/aminoacid.htm}.}\\
\multicolumn{4}{l}{$^\mathrm{c}$Silva and Annamalai, \textit{J. Thermodyn.} 2009, 186723 (2009)}\\
\multicolumn{4}{l}{$^\mathrm{d}$Roels, \textit{Energetics and kinetics in biotechnology}, Elsevier Biomedical Press (1983)}\\
\hline
\end{tabular}
\label{tab:combustion}
\end{table}

\subsection{A dynamical systems model of ontogeny based on DEB theory}

Figure~4b of the main text presents the dynamical systems model of ontogenetic development in a symbolic form:
\begin{align}
\dd{\mathbf{M}}{t} = \dot{\mathbf{F}}\lr{\dot{\mathbf{J}}_\mathrm{E}}
& \hspace{0mm}
& \dot{\mathbf{J}}_\mathrm{E} = \dot{\mathbf{J}}_\mathrm{E}\lr{\mathbf{M}; \bm{\upbeta}; X, T}.
\end{align}
This form emphasises that the rate of change of the state variables, collected in the vector $\mathbf{M}$, is given by a vector function $\dot{\mathbf{F}}$ that depends on the reaction rates, collected in the vector $\dot{\mathbf{J}}_\mathrm{E}$, of the three macrochemical reactions: assimilation, growth, and dissipation. The reaction rates, in turn, are a function of the state variables, as well as a vector of phenotypic parameters $\bm{\upbeta}$ and environmental forcings such as food density $X$ and temperature $T$. In what follows, we unpack all these functional dependencies.

The vector of state variables is $\mathbf{M}^\mathrm{T} = \lr{M_\mathrm{X}, M_\mathrm{E}, M_\mathrm{V}, M_\mathrm{P}, M_\mathrm{H}}$, where in addition to the variables introduced previously, $M_\mathrm{H}$ denotes maturity. The vector function $\dot{\mathbf{F}}$ accordingly has five components, interpreted as the rates of: food ingestion, reserve buildup, structural growth, feces egestion, and maturation. These components have the form:
\begin{equation}
\dot{\mathbf{F}}=
\begin{pmatrix}
y_\mathrm{XE}\dot{J}_\mathrm{EA}\\
\dot{J}_\mathrm{EA}-\dot{J}_\mathrm{EG}-\dot{J}_\mathrm{ED}\\
y_\mathrm{VE}\dot{J}_\mathrm{EG}\\
y_\mathrm{PE}\dot{J}_\mathrm{EA}\\
\dot{J}_\mathrm{EH}
\end{pmatrix},
\end{equation}
where $\dot{J}_\mathrm{EH}$ is not a fundamentally new flow of substances, but rather a part of the dissipation flow $\dot{J}_\mathrm{ED}$ directed at maturation. Importantly, the appearance of $\dot{J}_\mathrm{EH}$ in the last equation highlights the need for transitioning from the realm of physics (e.g., the organism as a thermodynamical system dissipates) to the realm of biology (e.g., what constitutes organismal dissipation and how it depends on the organism's state). We next review the biological assumptions specific to DEB theory that enable such a transition.

Staying focused on the dissipation flow, it is assumed that this flow satisfies the needs of three types of metabolic processes: somatic (i.e., structural) maintenance $\dot{J}_\mathrm{EM}$, maturity maintenance $\dot{J}_\mathrm{EJ}$, and maturation (until the organism matures) $\dot{J}_\mathrm{EH}$. If a cell on average requires a certain amount of reserve material for maintenance per unit of time, then $n$ cells require $n$ times the same amount, which is why the somatic maintenance flow is taken to be proportional to the amount of structure, \smash{$\dot{J}_\mathrm{EM} = \frac{\dot{k}_\mathrm{M}}{y_\mathrm{VE}}M_\mathrm{V}$}, where $\dot{k}_\mathrm{M}$ is a parameter called the somatic-maintenance rate coefficient (dimension: time$^{-1}$). Analogously, the maturity-maintenance flow is simply $\dot{J}_\mathrm{EJ} = \dot{k}_\mathrm{J}M_\mathrm{H}$, where $\dot{k}_\mathrm{J}$ is a parameter called the maturity-maintenance rate coefficient.

The maturation flow is a relatively complex function of the state variables; there is little value in writing down this functional dependence explicitly. To arrive at an instructive equation for the maturation flow, it is first assumed that mobilised reserve $\dot{J}_\mathrm{EG}+\dot{J}_\mathrm{ED}$ is split into a fraction $\kappa$ for somatic needs and the remaining fraction $1-\kappa$ for maturation needs. However, maturation can proceed only if this remaining fraction is sufficient to satisfy maturity maintenance. Also, maturation stops once the organism reaches adulthood. We therefore have:
\begin{equation}
\dot{J}_\mathrm{EH} = 
\begin{cases}
\lr{1-\kappa}\lr{\dot{J}_\mathrm{EG}+\dot{J}_\mathrm{ED}} - \dot{J}_\mathrm{EJ}, & M_\mathrm{H} < M_\mathrm{Hp}\\
0, & M_\mathrm{H} \geq M_\mathrm{Hp}
\end{cases},
\end{equation}
where the parameter marking sexual maturation $M_\mathrm{Hp}$ is called maturity at puberty.

Shifting attention to the growth flow, which is the other reserve-draining flow beside dissipation, we notice some symmetries with the maturation flow. Specifically, growth can proceed only if the fraction $\kappa$ of mobilised reserve $\dot{J}_\mathrm{EG}+\dot{J}_\mathrm{ED}$ is sufficient to satisfy somatic maintenance. We therefore have:
\begin{equation}
\dot{J}_\mathrm{EG} = \kappa\lr{\dot{J}_\mathrm{EG}+\dot{J}_\mathrm{ED}} - \dot{J}_\mathrm{EM}.
\end{equation}
This equation also leads to a relatively complex function of the state variables, but one that is instructive to write down explicitly:
\begin{equation}
\dot{J}_\mathrm{EG} = \frac{g \kappa y_\mathrm{VE} \frac{M_\mathrm{E}}{M_\mathrm{V}}\lr{\frac{M_\mathrm{Vm}}{M_\mathrm{V}}}^\frac{1}{3} - 1}{1+\kappa y_\mathrm{VE}\frac{M_\mathrm{E}}{M_\mathrm{V}}}\dot{J}_\mathrm{EM},
\label{eq:growthflow}
\end{equation}
where $g$ is a parameter called energy investment ratio. Together with $\dot{k}_\mathrm{M}$, this parameter measures how fast an organism drains the available reserve. The other parameter appearing in Eq.~\eqref{eq:growthflow} is the maximum amount of structural macromolecules $M_\mathrm{Vm}$ that the organism can accumulate. That such a maximum exists is not immediately visible, but we will make an intuitive argument to this effect next, and prove it rigorously later.

Eq.~\eqref{eq:growthflow} is the first to prominently feature the ratio of reserve to structure, \smash{$m_\mathrm{E}\defeq\frac{M_\mathrm{E}}{M_\mathrm{V}}$}, called reserve density. In the main text, we explained why reserve and structure should never outgrow one another, nor should they wildly fluctuate relative to one another. Under a stable feeding regimen, in fact, reserve density should relatively quickly stabilise. Assuming this to be true for the moment, the first term in the numerator of Eq.~\eqref{eq:growthflow} clearly decreases as structure grows, making sure that the whole numerator hits zero (marking the ceasing of growth) at a certain ultimate size $M_\mathrm{V\infty}$. At unlimited food abundance, moreover, the ultimate size converges to the maximum size, $M_\mathrm{V\infty}\rightarrow M_\mathrm{Vm}$, at which point the maximum reserve density equals
\begin{equation}
m_\mathrm{Em}\defeq\frac{M_\mathrm{Em}}{M_\mathrm{Vm}} = \frac{1}{g\kappa y_\mathrm{VE}}.
\end{equation}
If food abundance is limited, growth ceases when the ultimate size as a fraction of the maximum size equals
\begin{equation}
\frac{M_\mathrm{V\infty}}{M_\mathrm{Vm}} = \lr{g\kappa y_\mathrm{VE}\frac{M_\mathrm{E\infty}}{M_\mathrm{V\infty}}}^3 = \lr{\frac{m_\mathrm{E\infty}}{m_\mathrm{Em}}}^3,
\end{equation}
where we have used the definition \smash{$m_\mathrm{E\infty}\defeq\frac{M_\mathrm{E\infty}}{M_\mathrm{V\infty}}$}. The growth flow can now be simplified to:
\begin{equation}
\dot{J}_\mathrm{EG} = \frac{g \frac{m_\mathrm{E}}{m_\mathrm{Em}}\lr{\frac{M_\mathrm{Vm}}{M_\mathrm{V}}}^\frac{1}{3} - 1}{g+\frac{m_\mathrm{E}}{m_\mathrm{Em}}}\dot{J}_\mathrm{EM},
\end{equation}
where it is of note that the first term in the numerator scales with \smash{$M_\mathrm{V}^{-\frac{1}{3}}$}, but the somatic maintenance flow scales with \smash{$M_\mathrm{V}$}; multiplying these results in scaling with \smash{$M_\mathrm{V}^\frac{2}{3}$}. Such scaling reflects the fact that the reserve material for growth and dissipation is mobilised through the surface-area interface between reserve and structure.

The last flow that needs to be defined in terms of the state variables is the assimilation flow. This flow, like mobilisation for growth and dissipation, takes place through the surface-area interface between reserve and structure, which again suggests scaling with \smash{$M_\mathrm{V}^\frac{2}{3}$}. Noting that there is no feeding in the embryo stage, we have that
\begin{equation}
\dot{J}_\mathrm{EA} = 
\begin{cases}
0, & M_\mathrm{H} < M_\mathrm{Hb}\\
\frac{\dot{k}_\mathrm{M}M_\mathrm{Vm}^\frac{1}{3}}{\kappa y_\mathrm{VE}} f M_\mathrm{V}^\frac{2}{3}, & M_\mathrm{H} \geq M_\mathrm{Hb}\\
\end{cases},
\end{equation}
where $0\leq f\leq1$ is a fraction by which assimilation decreases at limited (relative to unlimited) food abundance, and $M_\mathrm{Hb}$ is maturity at birth.

We are now left with the task of proving that a constant $f$ leads to stable reserve density post birth. The dynamics of reserve density is
\begin{equation}
\dd{m_\mathrm{E}}{t} = \dd{}{t}\frac{M_\mathrm{E}}{M_\mathrm{V}} = \frac{1}{M_\mathrm{V}}\lr{\dd{M_\mathrm{E}}{t} - \frac{M_\mathrm{E}}{M_\mathrm{V}}\dd{M_\mathrm{V}}{t}}.
\end{equation}
From the last equation it follows that
\begin{equation}
\left.\dd{m_\mathrm{E}}{t}\right|_{m_\mathrm{E}=m_\mathrm{E\infty}} = 0 \Leftrightarrow \dd{M_\mathrm{E}}{t} = m_\mathrm{E\infty}\dd{M_\mathrm{V}}{t},
\end{equation}
or equivalently
\begin{equation}
\dot{J}_\mathrm{EA} - \dot{J}_\mathrm{EG} - \dot{J}_\mathrm{ED} = y_\mathrm{VE}m_\mathrm{E\infty}\dot{J}_\mathrm{EG}.
\label{eq:balance}
\end{equation}
After some algebra, we arrive at the following condition:
\begin{equation}
\frac{m_\mathrm{E\infty}}{m_\mathrm{Em}} = f,
\end{equation}
showing that $m_\mathrm{E\infty}=f m_\mathrm{Em}$ is an equilibrium point for reserve density. But is it stable? To answer this question, we rewrite Eq.~\eqref{eq:balance} as
\begin{equation}
\dot{J}_\mathrm{EA} - \dot{J}_\mathrm{ED} = \lr{1 + y_\mathrm{VE}m_\mathrm{E\infty}}\dot{J}_\mathrm{EG}.
\end{equation}
Only the right-hand side is dependent on reserve density; assuming that momentarily $m_\mathrm{E}<m_\mathrm{E\infty}$ leads to 
\begin{equation}
\dot{J}_\mathrm{EA} - \dot{J}_\mathrm{ED} > \lr{1 + y_\mathrm{VE}m_\mathrm{E\infty}}\dot{J}_\mathrm{EG} \Leftrightarrow \dd{m_\mathrm{E}}{t}>0,
\end{equation}
and $m_\mathrm{E}$ will increase towards $m_\mathrm{E\infty}$. Similar reasoning shows that assuming $m_\mathrm{E}>m_\mathrm{E\infty}$ leads to $m_\mathrm{E}$ decreasing towards $m_\mathrm{E\infty}$. The equilibrium point $m_\mathrm{E\infty}=f m_\mathrm{Em}$ is indeed stable.